\documentclass[%
reprint,
superscriptaddress,
amsmath,amssymb,
aps,
]{revtex4-1}

\usepackage{color}
\usepackage{graphicx}
\usepackage{dcolumn}
\usepackage{bm}

\begin {document}

\title{Infinite ergodic theory for three heterogeneous stochastic models with application to 
 subrecoil laser cooling }

\author{Takuma Akimoto}
\email{takuma@rs.tus.ac.jp}
\affiliation{%
  Department of Physics, Tokyo University of Science, Noda, Chiba 278-8510, Japan
}%

\author{Eli Barkai}
\affiliation{%
  Department of Physics, Bar-Ilan University, Ramat-Gan 5290002, Israel
}%

\author{G\"unter Radons}
\affiliation{%
  Institute of Physics, Chemnitz University of Technology, 09107 Chemnitz, Germany
}%


\date{\today}

\begin{abstract}
We compare ergodic properties of the kinetic energy for three stochastic models of subrecoil-laser-cooled gases. One model is based on a heterogeneous random walk (HRW), another is an HRW with long-range jumps 
(the exponential model), and the other is a mean-field-like approximation of the exponential model (the deterministic model).
All the models show an accumulation of the momentum at zero in the long-time limit, and 
a formal steady state cannot be normalized, i.e., there exists an infinite invariant density. 
We obtain the exact form of the infinite invariant density and the scaling function for the exponential and deterministic models
and devise a useful approximation for the momentum distribution in the HRW model. 
While the models are kinetically non-identical, it is natural to wonder whether their ergodic properties share common traits, 
given that they are all described by an infinite invariant density. We show that the answer to this question depends on the type of observable under study.
If the observable is integrable, the ergodic properties such as the statistical behavior of the time averages are universal as they are described by the Darling-Kac theorem.
In contrast, for non-integrable observables, the models in general exhibit non-identical statistical laws. This implies that 
focusing on non-integrable observables, we discover non-universal features of the cooling process, that hopefully can lead to a better understanding of the particular model most suitable for a statistical description of the process. 
This result is expected to hold true for many other systems, beyond laser cooling.
\end{abstract}

\maketitle


\section{Introduction}
In many cases in equilibrium statistical physics, a steady-state solution of a master equation yields the 
equilibrium distribution. However, the formal steady-state solution may not be normalizable, 
 especially for non-stationary stochastic processes found in the context of anomalous diffusion and 
non-normalizable Boltzmann states \cite{van1992stochastic, Kessler2010,lutz2013, Rebenshtok2014, Holz2015, Leibovich2019,Aghion2019,aghion2020infinite,aghion2021moses,Streissnin2021}. 
 Such an unnormalized formal steady state is called an infinite invariant density, 
 which is known from deterministic dynamical systems \cite{Thaler1983,Aaronson1997}. 
Interestingly, dynamical systems with infinite invariant densities exhibit non-stationary behaviors and 
trajectory-to-trajectory fluctuations of time averages, 
whereas they are ergodic in the mathematical sense \cite{Aaronson1997}.

The ergodic properties of dynamical systems with infinite invariant densities 
have been established in infinite ergodic theory \cite{Aaronson1997, inoue1997ratio, Thaler1998,Thaler2002, inoue2004ergodic, Akimoto2008,Akimoto2015,Sera2019,Sera2020}, where distributional limit theorems for time-averaged quantities play an important role. 
The distributional limit theorems state that time-averaged observables obtained with single trajectories show 
trajectory-to-trajectory fluctuations. The distribution function of the fluctuations depends on whether 
the observable is integrable with respect to the infinite invariant measure \cite{Aaronson1981,Akimoto2008,Akimoto2010,Akimoto2012,Akimoto2015}. This distributional behavior of time averages 
is a characteristic feature of infinite ergodic theory. Similar distributional behaviors have been 
 observed in experiments such as the fluorescence of quantum dots, diffusion in living cells, and interface fluctuations in liquid crystals 
 \cite{Brok2003,stefani2009,Golding2006,Weigel2011,Jeon2011,Hofling2013,Manzo2015,takeuchi2016}.

Subrecoil laser cooling is a powerful technique for cooling atoms \cite{cohen1990new, Bardou1994}. A key idea of this technique is to realize 
experimentally a heterogeneous random walk (HRW) of the atoms in momentum space. 
In a standard cooling technique such as Doppler cooling, a biased random walk is utilized to shift the momenta of
 atoms towards zero \cite{cohen1990new}. Thus, Doppler cooling is routinely modeled using a standard 
 Fokker--Planck equation for the momentum distribution. 
In contrast to a homogeneous random walk, an HRW 
enables the accumulation of walkers at some point without an external force induced by the Doppler effect. 
In other words,  the probability of finding a random walker 
at that point converges to one in the long-time limit
 due to an ingenious trapping mechanism, that gives rise to a heterogeneous environment. Hence,  
for subrecoil laser cooling, 
instead of a biased random walk, an HRW 
plays an essential role. This was a paradigm shift for cooling and useful for cooling beyond the lowest 
limit obtained previously in standard cooling techniques \cite{cohen1990new}.  

It has been recognized that infinite ergodic theory provides a fundamental theory for subrecoil-laser cooling \cite{Barkai2021,*barkai2022gas}. 
In \cite{Bardou2002} three models of subrecoil laser cooling are proposed. One is based on the 
HRW, another is obtained from the HRW model with long-range jumps
 called the exponential model, and the third is a mean-field-like approximation of the exponential model called 
the deterministic model. It is known that the infinite invariant density depends in principle on some details of the system \cite{Aghion2019,aghion2020infinite,aghion2021moses}. 
The question then remains: what elements of the infinite ergodic theory remain universal? 
 These questions with respect to the general validity of the theory are particularly important because we have at least two general classes of observables, i.e., integrable and non-integrable with respect to the infinite invariant measure. 
 To unravel the universal features of subrecoil laser cooling,  we explore here the three models of subrecoil laser cooling. 

 The rest of the paper is organized as follows. In Sec.~II, we introduce the three stochastic models of subrecoil laser cooling. 
 In Sec.~III, we introduce the master equation and the formal steady-state solution, i.e., the infinite invariant density, in the HRW model. 
 In Sections~IV and V, we present the infinite invariant densities and the distributional limit theorems for the time average of the kinetic energy in the deterministic and exponential model, respectively. While the master equations for the HRW and exponential model are different, we show that the propagators and the distributional behaviors of the time-averaged kinetic energy match very well. 
Section VI is devoted to the conclusion. In the Appendix, we give a derivation of the moments of the associated action as a function of 
time $t$.

\section{Three stochastic models}
   
   Here, we introduce the three stochastic models of subrecoil laser cooling. All the models 
   describe stochastic dynamics of the momentum of an atom. 
   
\begin{figure}
\includegraphics[width=.95\linewidth, angle=0]{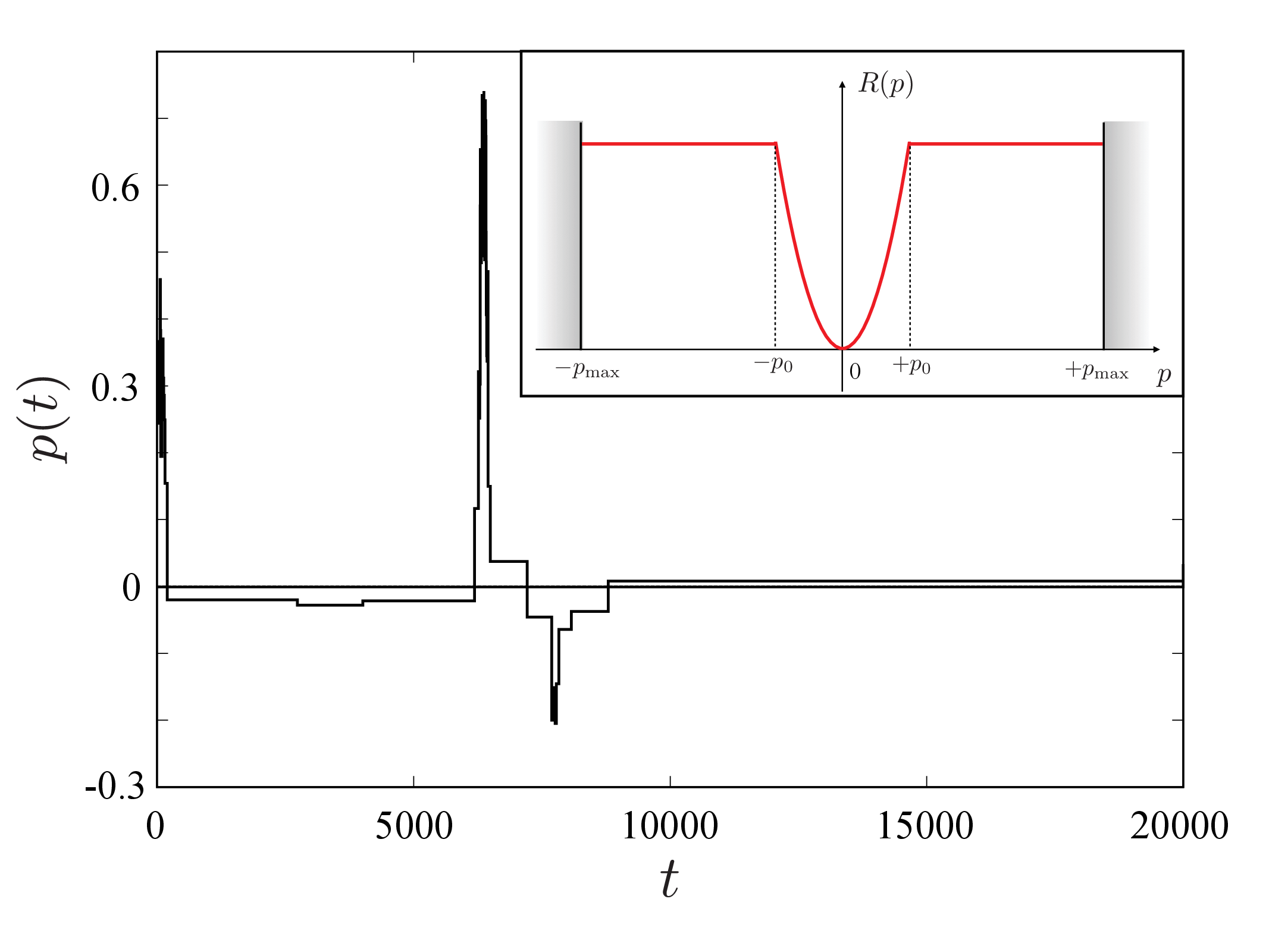}
\caption{A typical trajectory of momentum $p(t)$ in the HRW model, where $R(p)=|p|^2$, $p_0=1$, $p_{\max}=3$, and $\sigma^2=0.01$.
The inset is a schematic illustration of the jump rate $R(p)$.}
\label{traj}
\end{figure}

   First, the HRW model is a one-dimensional continuous-time random walk (CTRW)  in momentum space  $p$.  
   Here, we consider confinement, which is represented by reflecting boundary at $p=-p_{\max}$ and $p_{\max}$.
   The CTRW is a random walk with continuous waiting times. Usually, in the CTRW the waiting times are 
   independent and identically distributed (IID). In the HRW model, they are not  IID random variables. 
   In the HRW, 
   the waiting time between stochastic updates of momentum given $p$ is exponentially distributed with a mean waiting time $1/R(p)$. 
   After waiting the atom jolts and momentum is modified. 
   We assume that the jump distribution $G( \Delta p)$ follows a Gaussian distribution:
   \begin{equation}
G( \Delta p)=(2\pi \sigma ^{2})^{-1/2}\exp [-\Delta p^{2}/(2\sigma ^{2})],  
\end{equation}
where $\Delta p$ is a jump of the 
momentum of an atom and $\sigma^2$ is the variance of the jumps.  
   The heterogeneous rate $R(p)$ is important to cool atoms and can be realized by velocity selective coherent population trapping in experiments \cite{AAK88}. In subrecoil laser cooling, the jump rate $R(p)$ is typically given by 
\begin{equation}
R(p)\propto |p|^\alpha 
\label{atom-laser interaction}
\end{equation}
for $|p|\to 0$ \cite{Bardou2002}, where $\alpha$ is a positive constant. 
This constant can take any value in principle \cite{KaC92}, for instance,
 $\alpha=2$ in velocity-selective coherent population trapping \cite{AAK88}. 
In what follows, we consider a specific jump rate: 
\begin{equation}
R(p)= \left\{
\begin{array}{ll}
c^{-1}|p|^\alpha  \quad &(|p| <p_0)\\
\\
c^{-1}|p_0|^\alpha &(|p| \geq p_0),
\end{array}
\right.
\label{atom-laser interaction2}
\end{equation}
where $p_0$ is the width of the jump rate dip and $c$ is a positive constant (see Fig.~\ref{traj}). 
At $p=\pm  p_{\max}$ we have reflecting boundary. 
A typical trajectory in the HRW model is shown in Fig.~\ref{traj}. 
   Since the HRW model is a non-biased random walk, the momentum will eventually reach high values.
To prevent such a situation, one considers a confinement in an experimentally realizable way. 

Next, we explain how we obtain the other two models, i.e., the exponential and the deterministic model, inspired by the HRW model. 
The region in momentum space can be divided into two regions, i.e., 
trapping and recycling regions \cite{Bardou2002}. 
The trapping region is defined as $|p| \leq p_{\rm trap}$, where we assume $p_{\rm trap} \ll \sigma$ and $p_{\rm trap} < p_0$. The assumption  
$p_{\rm trap} \ll \sigma$ is used in the uniform approximation stated below. 
In the recycling region, the atom undergoes a non-biased random walk, which will eventually lead the atom back to 
the trapping region with the aid of the confinement. 
The jumps of a random walker are long-ranged in the trapping region in the sense that 
momentum after jumping in the trapping region is approximately independent of the previous momentum. 
Therefore, the following assumption is quite reasonable. In the exponential and the deterministic model, 
 momentum after jumping in the trapping region is assumed to be an IID random variable. 
 In particular, the probability density function (PDF) $\chi (p)$ for the momentum at every jump in the trapping region 
 is assumed to be uniform \cite{Saubamea1999,Bardou2002,bertin2008}: 
\begin{equation}
\chi  (p)= \frac{1}{2p_{\rm trap}} ~~{\rm for}~p\in[-p_{\rm trap},p_{\rm trap}].
\label{chi}
\end{equation}
A trajectory for the exponential model is similar to that for the HRW model.
However, a crucial difference between the HRW model and the exponential model is in the nature of the waiting time: the waiting time is an independent random variable in the exponential model, whereas it is not in the HRW model.  
In the HRW model, momentum performs a random walk. When momentum changes due to photon scattering, the renewed momentum depends on the previous
 momentum. Hence in this sense we have a correlation of momentum that spans several jolting events. 
 On the other hand, the renewed momentum is independent of the previous momentum in the exponential model. 
 In both models, the waiting time given $p$ is an exponentially distributed random variable with rate $R(p)$. 
 Thus, the statistics of the waiting times in the two models is different, because in the HRW model they are correlated through the momentum sequence, 
 whereas in the exponential model they are not.
 However, in the exponential model, the momentum is always in the trapping region. In the HRW model, it jumps in the recycling region. In other words, 
 a time of returning to the trapping region is not taken into consideration in the exponential model. 


A difference between the exponential and the deterministic model is in the coupling between the waiting time and the momentum. 
In the exponential model, momentum and waiting time are stochastically coupled. 
As for the HRW this model is a Markov model and the conditional PDF of the waiting time given the momentum $p$ follows an exponential 
distribution with mean $1/R(p)$. 
On the other hand, the deterministic model is a non-Markov model.  
The waiting time given the momentum $p$ is deterministically prescribed as $\tau (p) = 1/R(p)$ \cite{Bardou2002}. In other words, the waiting time, which is 
a random variable in the exponential model, is replaced by its mean in the deterministic model. In this sense, the deterministic 
model is a mean-field-like model of the exponential model.
Note that this implies a double meaning of $1/R(p)$: while in the HRW and in the exponential model 
it is the mean waiting time, whereas in the deterministic model it is the exact waiting time for a given momentum $p$.

\section{Heterogeneous Random Walk Model}

Here, we consider the HRW model confined to the interval $[-p_{\max},p_{\max}]$ \cite{AAK88,Bardou1994}. 
The momentum $p(t)$ at time $t$ undergoes a non-biased random walk. 
 Jumps of the momentum are attributed to photon scattering and spontaneous emissions.  
 Importantly, its jump rate $R(p)$ follows Eq.~(\ref{atom-laser interaction}) for $|p|<p_0$ \cite{Bardou1994}. 
In this model, the conditional PDF $q(\tilde{\tau}|p)$ of $\tilde{\tau}$ given $p$ follows the exponential distribution:
\begin{equation}
q(\tilde{\tau}|p)= R(p) \exp(-R(p) \tilde{\tau}).
\label{conditional_PDF_HRW}
\end{equation}
Clearly, the mean waiting time given $p$ explicitly depends on $p$ when $|p| <p_0$.
Thus, the random walk is heterogeneous. 
A confinement of atoms can also be achieved by Doppler cooling \cite{cohen1990new, Bardou1994}. 
However,  for simplicity, we consider reflecting boundary conditions at $p=\pm p_{\max}$. 
As will be observed later, the size of the confinement or the width of the jump rate dip 
does not affect the asymptotic behavior of the scaling function of the propagator. 
More precisely, the scaling function and fluctuations of the time-averaged energy 
do not depend on $p_{\max}$ and $p_0$. 
As shown in Fig.~\ref{traj}, the momentum of an atom remains 
constant for a long time when $|p|$ is small. On the other hand, 
momentum changes frequently occur when $|p|$ is away from zero.

\subsection{Master equation and infinite invariant density}
The HRW model is a Markov model. In general, the time evolution of the propagator of a Markov model can be described by a master equation \cite{van1992stochastic}. 
The time evolution of the probability density function (PDF) $\rho(p,t)$ of  momentum $p$ at time $t$ 
is given by the master equation with gain and loss terms:
\begin{equation}
\frac{\partial \rho \left( p,t\right) }{\partial t}=\int_{-p_{\max}}^{p_{\max}} dp^{\prime }\left[ W(p^{\prime}\rightarrow p)
\rho \left( p^{\prime },t\right) -W(p\rightarrow p^{\prime})\rho \left( p,t\right) \right],  
\label{Master}
\end{equation}
where $W(p\rightarrow p^{\prime})$ is the transition rate from $p$ to $p'$. As will be shown later, the formal steady-state solution for 
the master equation may not provide a PDF but a non-normalized density, i.e. an infinite invariant density.
Jump and transition rates can be represented as 
\begin{equation}
R(p)=\int_{-\infty}^\infty dp^{\prime }W(p\rightarrow p^{\prime })
\label{jump rate}
\end{equation}%
and
\begin{equation}
W(p\rightarrow p^{\prime })= R(p) \tilde{G}(p'|p), 
\label{transition rate0}
\end{equation}
respectively, where $ \tilde{G}(p'|p)$ is the conditional PDF of $p'$ given $p$, 
 where both the domain and the codomain of the function $ \tilde{G}(p'|p)$ are  $[-p_{\max}, p_{\max}]$ because of the confinement.  
The function $ \tilde{G}(p'|p)$ is equivalent to $G(p'-p)$ when $p+\Delta p$ does not exceed the boundary, i.e., $|p+\Delta p| < p_{\max}$, where $\Delta p$ is 
a momentum jump following the Gaussian distribution.
On the other hand, $ \tilde{G}(p'|p)$ cannot depend solely on the 
difference $p'-p$ when a random walker reaches the reflecting boundary, i.e., $|p+\Delta p| > p_{\max}$.   
In particular, we have 
\begin{equation}
\tilde{G}(p^{\prime }|p)= \sum_{n=-\infty}^\infty G(2np_{\max } -p  + (-1)^{n}p^{\prime}).
\end{equation}
Because $G(x)$ is a symmetric function (Gaussian distribution), $\tilde{G}(p^{\prime }|p)$ is symmetric in $p$ and $p'$: $\tilde{G}(p^{\prime }|p)=\tilde{G}(p|p^{\prime })$.
It follows that the master equation  (Eq.~(\ref{Master})) of the HRW model takes the following form:%
\begin{equation}
\frac{\partial \rho \left( p,t\right) }{\partial t}=-R(p)\rho \left( p,t\right) +\int_{-p_{\max}}^{p_{\max}}
dp^{\prime }\rho \left( p^{\prime },t\right) R(p^{\prime }) \tilde{G}(p|p^{\prime }).
\label{Master1}
\end{equation}

The stationary solution $\rho^*(p)$ is easily obtained from the detailed balance in Eq.~(\ref{Master}), i.e., 
\begin{equation}
 W(p^{\prime }\rightarrow p)\rho ^{\ast }\left( p^{\prime }\right)
-W(p\rightarrow p^{\prime })\rho ^{\ast }\left( p\right)  =0,
\label{detailed balance}
\end{equation}%
where $\rho^*(p)$ is the stationary solution. 
As shown before, the conditional PDF $\tilde{G}(p|p')$ is symmetric, i.e., $\tilde{G}(p|p')=\tilde{G}(p'|p)$. Therefore, 
detailed balance yields 
\begin{equation}
R(p^{\prime })\rho ^{\ast }\left( p^{\prime }\right) =R(p)\rho ^{\ast}\left( p\right) ,
\end{equation}
which is fulfilled only if $R(p)\rho ^{\ast }\left( p\right) $ is constant. 
In subrecoil laser cooling, the jump rate $R(p)$ becomes a power-law form near $p\cong 0$, i.e., Eq.~(\ref{atom-laser interaction}). 
 For example, the velocity selective coherent population trapping gives $\alpha=2$ 
\cite{AAK88}, and the Raman cooling experiments realize  $\alpha=2$ and 4 by 1D square pulses and the 
Blackman pulses, respectively \cite{RBB95}.
Therefore,  for $|p| \ll p_{\max}$, the steady-state distribution $\rho^* (p)$ is formally given by
\begin{equation}
\rho^* (p) ={\rm const.}/R(p) \propto |p|^{-\alpha}.
\label{steady-state}
\end{equation}
For $\alpha\geq 1$, it cannot be normalized because of the divergence at $p=0$, and $\rho^* (p)$ 
is therefore called an infinite invariant density. Although $\rho^* (p)$ is 
the formal steady state, a steady state in the conventional sense does not exist in the system with $\alpha\geq 1$. 
  As will be shown below, a part of the infinite invariant density can be observed in the 
  propagator especially for a large time. Moreover, it will be shown that 
  $t^{1-1/\alpha} \rho (p,t)$ converges to the infinite invariant density for $t\to\infty$.
  Therefore, the infinite invariant density is not a vague solution but plays an important role in reality.

\begin{figure}
\includegraphics[width=.95\linewidth, angle=0]{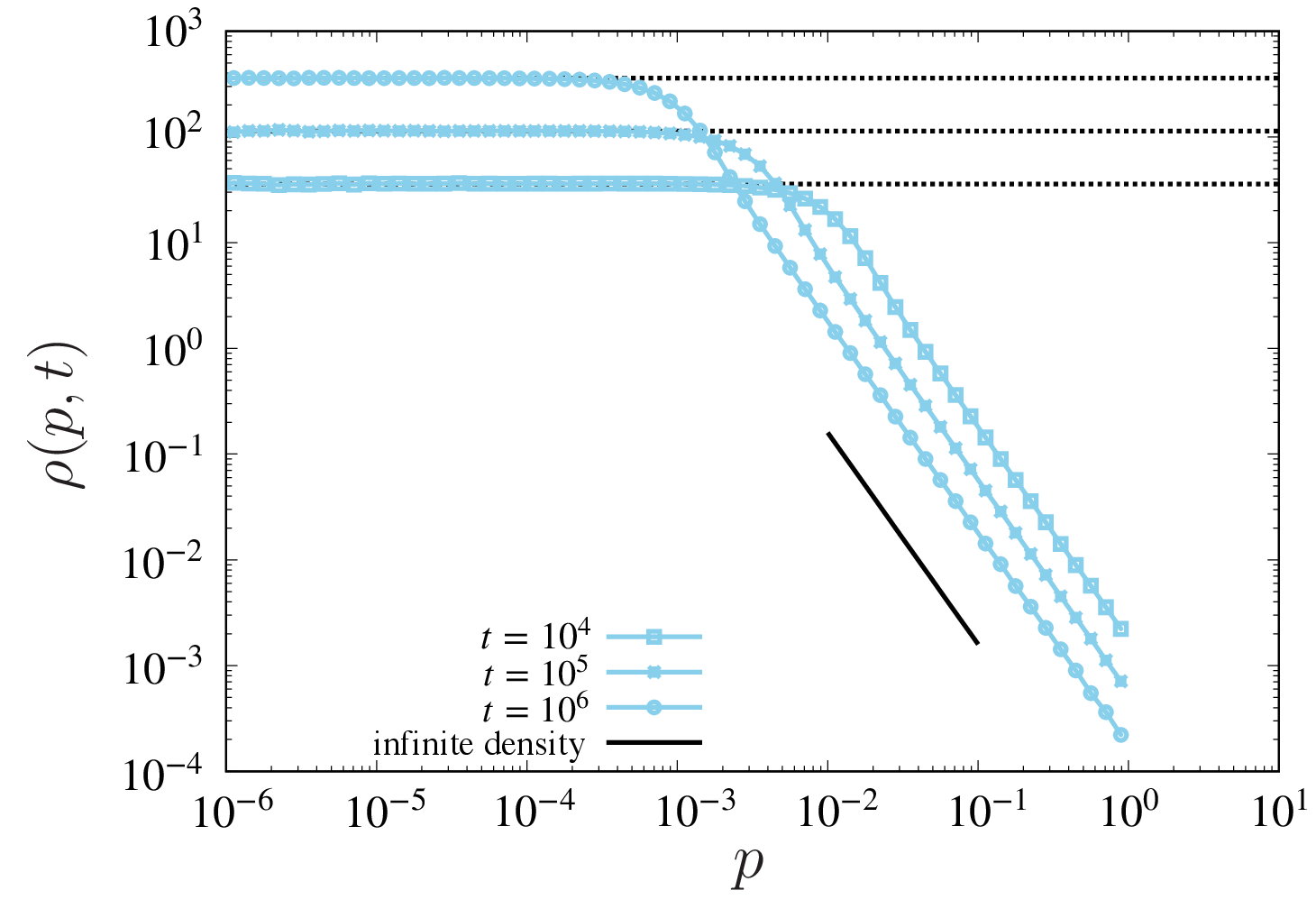}
\caption{Time evolution of the propagator in the HRW model ($p_0=p_{\max}=1$, $\sigma=1$, and $R(p)=|p|^2$). 
Symbols with lines are the numerical results of the HRW model by simulating trajectories of random walkers. 
The solid line represents a part of a steady-state solution,  $\rho^*(p) \propto |p|^{-\alpha}$, for reference. 
The dashed lines represent plateaus around $|p|= 0$, which shift up with time $t$. 
Initial momentum is chosen uniformly on $[-1,1]$.  The number of trajectories used in this and all subsequent simulation results is $10^6$. }
\label{prop-hrw}
\end{figure}

Figure~\ref{prop-hrw} shows numerical simulations of the propagator in the HRW model. 
The propagator accumulates 
near zero, and $\rho (p,t)$ around $p\cong 0$ increases with time $t$. Moreover, a power-law form, i.e., $p^{-\alpha}$, 
of the formal steady state $\rho^*(p)$ is observed, especially when $t$ is large, except for $p\cong 0$ (see also Fig.~\ref{propagator-exp}). 
Since the infinite invariant density $\rho^*(p)$ cannot be normalized, the propagator never converges to $\rho^*(p)$.

\section{Exponential model}
In this section, we give theoretical results for the exponential model, which were already shown in our previous study 
\cite{Barkai2021,*barkai2022gas}.
 Here, we consider the Laplace transform of the propagator and execute the inverse transform to obtain the infinite invariant 
 density and the scaling function. 
 The derivation of the scaling function is different from the previous study \cite{Barkai2021,*barkai2022gas}, where 
 the master equation is directly solved.

\subsection{Master equation, infinite invariant density, and scaling function}
In the exponential model, the jump distribution is independent of the previous momentum unlike the HRW model. 
Therefore, for the exponential model the conditional probability $\tilde{G}(p'|p)$ in Eq.~(\ref{transition rate0}) can be replaced by a $p$-independent function 
$\chi(p')$ leading to
 \begin{equation}
 W(p\rightarrow p^{\prime })= R(p)\chi (p^{\prime }).
\label{transition rate exp}
\end{equation}
Inserting this into Eq.~(\ref{Master}), the
 master equation of the exponential model becomes
\begin{equation}
\frac{\partial \rho \left( p,t\right) }{\partial t}= - R(p)\rho(p,t) + \frac{1}{2p_{\rm trap}} \int_{-p_{\rm trap}}^{p_{\rm trap}} R(p') \rho(p',t) dp',
\label{Master-exp}
\end{equation}
where we used Eq.~(\ref{chi}). As a result the second term, i.e., gain term, is different from that in the HRW model, Eq.~(\ref{Master1}). 
In the exponential model, the momentum remains constant until the next jump, and the conditional waiting time distribution given by momentum $p$  follows an exponential distribution with mean $1/R(p)$, which is the same as in the HRW model, i.e., Eq.~(\ref{conditional_PDF_HRW}) holds also here.  
 Because the conditional waiting time distribution depends on $p$, 
the joint PDF of momentum $p$ and waiting time $\tilde{\tau}$, 
\begin{equation}
\phi  (p,\tilde{\tau} )=\left\langle \delta \left( p-p_{i}\right) \delta \left( \tilde{\tau}-\tilde{\tau}_{i}\right) \right\rangle  
\label{jpdgen}
\end{equation}
 plays an important role, where $\delta \left(.\right) $ is the $\delta$ function, $\langle \cdot \rangle$ represents the ensemble average, 
 $i$ is the $i$-th emission  event ($i=1,2, ...$), $p_i$ is the $i$th momentum, and $\tilde{\tau}_{i}$ is the $i$th waiting time. It can be expressed by 
\begin{equation}
\phi  (p,\tilde{\tau} )= q(\tilde{\tau}|p) \chi (p) ,
\label{joint-pdf exp}
\end{equation} 
where $q(\tilde{\tau}|p)$ is the conditional PDF $q(\tilde{\tau}|p)$ of waiting time $\tilde{\tau}$ given $p$, Eq.(\ref{conditional_PDF_HRW}), 
and $\chi (p)$ is given by Eq.(\ref{chi})

The unconditioned PDF of the waiting time is given by 
\begin{eqnarray}
\psi (\tilde{\tau}) = \frac{1}{2p_{\rm trap}} \int_{-p_{\rm trap}}^{p_{\rm trap}} R(p) \exp (-R(p)\tilde{\tau})dp,
\end{eqnarray} 
which follows from averaging the joint PDF,
 over the uniform density $\chi(p)$. 
By a change of variables ($y=R(p)\tilde{\tau}$), we have
\begin{eqnarray}
\psi (\tilde{\tau}) &=& \frac{c^{\frac{1}{\alpha}}\tilde{\tau}^{-1-\frac{1}{\alpha}}}{\alpha p_{\rm trap}} \int_0^{\tilde{\tau} c^{-1} p_{\rm trap}^\alpha} 
y^{\frac{1}{\alpha}} \exp (-y)dy\\
&\sim& \frac{\gamma c^{\gamma}\Gamma (1+\gamma)}{ p_{\rm trap}} \tilde{\tau}^{-1-\gamma}\quad (\tilde{\tau}\to \infty),
\end{eqnarray} 
where $\gamma=1/\alpha$.
In what follows, we assume $\gamma\leq 1$, which implies that the mean waiting time 
diverges. Therefore, as will be shown, the dynamics of $p$ becomes non-stationary.  


The exponential model is a continuous-time Markov chain, which is a special type of semi-Markov process (SMP). Therefore, 
we utilize an SMP with continuous variables to obtain analytical results for the exponential model.
In an SMP, the state value is determined by the waiting time, which is randomly 
selected, or equivalently, the waiting time is determined by the state value, which is randomly chosen. 
In the latter case, the state value is renewed according to the PDF $\chi(p)$.   
In general, an SMP is characterized by the state distribution $\chi(p)$ and the joint PDF of the state value and the waiting time $\phi (p,\tau)$, Eq.~(\ref{joint-pdf exp}). 
The deterministic model, which we will treat in Sect.~V, is identical to the SMP with a deterministic coupling 
between the state value and the waiting time. 
On the other hand, the SMP with an exponential conditional PDF of waiting times given the state 
is equivalent to the exponential model. For the SMP with $\chi(p)$ and $\phi (p,\tau)$,  
the Laplace transform of the propagator with respect to $t$ 
is obtained as in Ref.~\cite{Akimoto2020}. Applying the technique given in Ref.~\cite{Akimoto2020} to the exponential model, we find 
\begin{equation}
\hat{\rho} (p,s) = \frac{1}{s} \frac{\chi(p) - \hat{\phi}(p,s)}{1-\hat{\psi}(s)},
\label{MW-SMP}
\end{equation}
where $\hat{\phi}(p,s)$ and $\hat{\psi}(s)$ are the Laplace transforms of $\phi(p,\tilde{\tau})$ and $\psi(\tilde{\tau})$ with respect to 
$\tilde{\tau}$, respectively. Here,  initial conditions as for ordinary renewal processes were used \cite{Akimoto2020,Cox1962}.

In the exponential model, the Laplace transform of the joint PDF is given by 
\begin{equation}
\hat{\phi}(p,s)= \frac{\chi (p) R(p)}{s+R(p)}.
\label{JPDF-SMP}
\end{equation}
If follows from Eqs.~(\ref{MW-SMP}) and (\ref{JPDF-SMP}) that $\hat{\rho} (p,s)$ becomes 
\begin{equation}
\hat{\rho} (p,s) =  \frac{ \chi(p)}{s+R(p)} \frac{1}{1-\hat{\psi}(s)}.
\end{equation}
In the long-time limit ($s\to 0$), it becomes 
\begin{equation}
\hat{\rho} (p,s) \cong  \frac{ 1}{s+c^{-1}|p|^{\alpha}} \frac{1}{2\Gamma(1-\alpha^{-1})\Gamma(1+\alpha^{-1}) (cs)^{\alpha^{-1}}},
\end{equation}
where 
$\chi(p)=1/(2p_{\rm trap})$ is used. Interestingly, the Laplace transform 
of the propagator does not depend on $p_{\rm trap}$ in the long-time limit.  To obtain the exponential model from the HRW model, we assumed that
 $p_{\rm trap}$ is much smaller than $\sigma$. However, the asymptotic  behavior of the propagator is independent of $p_{\rm trap}$ in the exponential model. 
 Therefore, $p_{\rm trap}$ introduced in the HRW model can be assumed to be arbitrary small because the value of $p_{\rm trap}$ does 
 not affect the asymptotic behavior of the propagator of the exponential model. 
 When $p_{\rm trap}\ll \sigma$, the distribution of momentum after jumping in the trapping region, i.e., $[-p_{\rm trap}, p_{\rm trap}]$, is approximately uniform.
 Therefore, 
the exponential model with the uniform approximation for $\chi(p)$ is a good approximation for the HRW model for large $t$. 
By the inverse Laplace transform, we have 
\begin{equation}
\rho(p,t)  \cong \frac{\sin (\pi \alpha^{-1})}{2\pi c^{\alpha^{-1}}  \Gamma (1+\alpha^{-1})} \int_0^t dt' e^{-c^{-1} |p|^{\alpha} (t-t')} t'^{\alpha^{-1} -1}
\label{propagator_asympt-exp}
\end{equation}
for $t\to\infty$. Through a change of variables ($u=t'/t$), we obtain 
\begin{equation}
\rho(p,t)  \cong \frac{\sin (\pi \alpha^{-1}) t^{\alpha^{-1} }}{2\pi c^{\alpha^{-1}} \Gamma (1+\alpha^{-1})} \int_0^1 du 
e^{-c^{-1} |p|^{\alpha} t(1-u)} u^{\alpha^{-1} -1}.
\label{propagator_asympt-exp2}
\end{equation}
Therefore, the cooled peak, i.e., $\rho(0,t)$, increases with $t^{\alpha^{-1}}$, which means that the probability 
of finding the cooled state ($p\cong 0$) increases with time, i.e., this is a signature of cooling.

For $|p|>0$ and $t\gg 1$, the integral in Eq.~(\ref{propagator_asympt-exp2}) can be approximated leading to
\begin{equation}
\rho(p,t)  \cong \frac{\sin (\pi \alpha^{-1}) t^{\alpha^{-1} -1}}{2\pi c^{\alpha^{-1}-1}  \Gamma (1+\alpha^{-1})} 
 \frac{1}{|p|^{\alpha} }.
\label{propagator_asympt-exp3}
\end{equation}
Furthermore, an infinite invariant density is obtained as 
\begin{equation}
  \lim_{t\to \infty} t^{1-\alpha^{-1}} \rho(p,t) = I_{\rm exp} (p) \equiv
 \frac{ \sin (\pi \alpha^{-1} ) \left\vert p\right\vert ^{-{\alpha}}}{2 \pi c^{\alpha^{-1}-1} \Gamma (1+\alpha^{-1}) }
\label{inf-d-exp}
\end{equation}
for $|p| \leq p_{\rm trap}$. The power-law form of Eq.~(\ref{inf-d-exp}), 
$I_{\rm exp} (p) \propto |p|^{-\alpha}$, in the exponential model matches with 
the infinite invariant density, Eq.~(\ref{steady-state}), in the HRW model.

Through a change of variables ($p'=t^{\alpha^{-1}} p/c^{\alpha^{-1}}$),
we obtain the rescaled propagator $\rho_{\rm res} (p',t)$. In the long-time 
limit, the rescaled propagator converges to a time-independent function $g_{\rm exp} (p')$ (scaling function):
\begin{equation}
 \rho_{\rm res} (p',t) \equiv \rho (c^{\alpha^{-1}} p'/t^{\alpha^{-1}},t) \left| \frac{dp}{dp'}\right|
\to g_{\rm exp} (p')  ,
\label{rescaling}
\end{equation}
where the scaling function is given by
\begin{equation}
 g_{\rm exp} (p') \equiv  \frac{\sin (\pi \alpha^{-1}) }{2\pi  \Gamma (1+\alpha^{-1})} \int_0^1 du 
e^{- |p'|^\alpha (1-u)} u^{\alpha^{-1} -1}.
 \label{sf-exp}
\end{equation}
This scaling function describes the propagator near $p=0$. 
This result was previously obtained by a different approach \cite{Barkai2021,*barkai2022gas}. 

\begin{figure}
\includegraphics[width=.95\linewidth, angle=0]{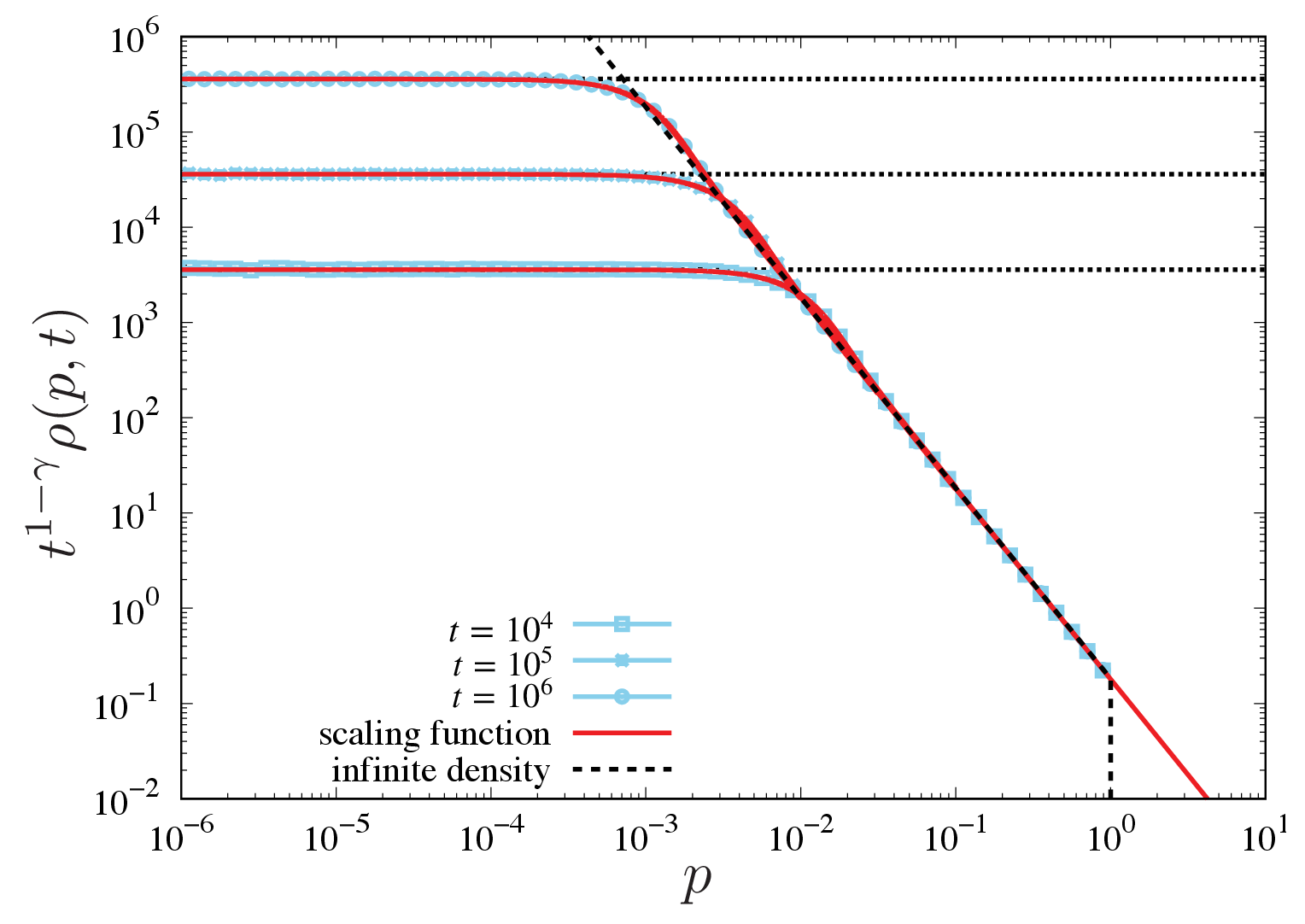}
\caption{Time evolution of the propagator, i.e. data from Fig.~\ref{prop-hrw}, multiplied by $t^{1-\gamma}$ in the HRW model 
for different times ($\alpha =2$,  $c=1$, $p_0=p_{\max}=1$, and $\sigma^2 =1$). 
Symbols with lines are the results of numerical simulations for the HRW model. 
The dashed lines represent the infinite invariant density, i.e., Eq.~(\ref{inf-d-exp}). The solid lines represent rescaled scaling functions, 
$t g_{\rm exp} (t^\gamma p)$. The dotted lines represent $t g_{\rm exp} (0)$ for different values of $t$. 
The initial momentum is chosen uniformly on $[-1,1]$.  
 }
\label{propagator-exp}
\end{figure}

\begin{figure}
\includegraphics[width=.95\linewidth, angle=0]{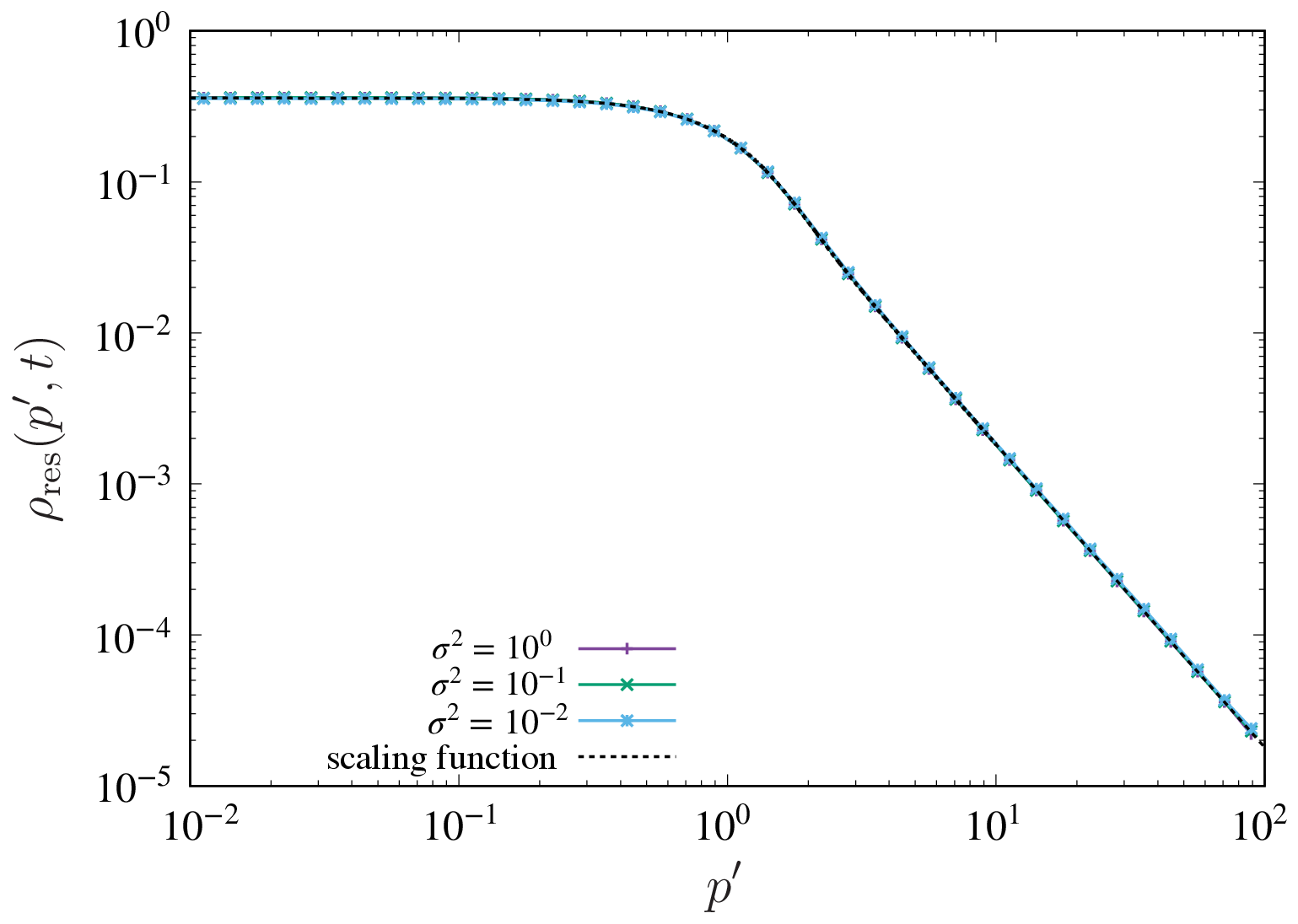}
\caption{Rescaled propagator of the HRW model for different values of $\sigma^2$  ($\alpha =2$, $c=1$, $p_0=p_{\max}=1$, and $t=10^4$).
Symbols with lines are the results of numerical simulations for the HRW model. 
The dashed solid line represents the scaling function, i.e., Eq.~(\ref{sf-exp}). 
The initial position is chosen uniformly on $[-1,1]$. 
Note that the results for different $\sigma^2$ are indistinguishable.}
\label{propagator-rescale-exp}
\end{figure}

Here, we are going to demonstrate 
 that the theory of the exponential model describes the asymptotic behavior of the propagator 
in the HRW model surprisingly well.  
Figure~\ref{propagator-exp} shows that the propagator for the HRW model 
is in perfect agreement with the analytical result of the exponential model, i.e., Eq.~(\ref{propagator_asympt-exp2}). 
In the numerical simulations of the HRW model, we generated $10^8$ trajectories to obtain the propagator. 
There are two forms in the propagator. 
The propagator near $p=0$ increases with 
time $t$. On the other hand, the propagator for $p>0$ asymptotically approaches a power-law form, i.e., the infinite invariant density. 
Figure~\ref{propagator-rescale-exp} shows that the rescaled propagator of the HRW model for different times 
is well captured by the scaling function $g_{\rm exp} (p')$ without fitting parameters, where we generated $10^8$ trajectories 
to obtain the rescaled propagator. 
 Because the scaling function 
describes the details of the propagator near $p=0$ and is universal in the sense that it does not depend on $p_{\rm trap}$ in the exponential 
model, 
the dynamics of the HRW model near $p=0$ should also be universal and does not depend on the details of the jump 
distribution $G(\Delta p)$. In fact, as shown in Fig.~\ref{propagator-rescale-exp}, 
the rescaled propagator does not depend on $\sigma^2$. 
This is one of the reasons why the uniform approximation works very well. 
Moreover, because the momentum almost certainly approaches zero in the long-time limit, the assumption of $|p|\ll 1$ is correct for $t\gg 1$. 
Furthermore, it can be confirmed that Eq.~(\ref{propagator_asympt-exp2}) becomes a solution to the master equation, Eq.~(\ref{Master1}),
in the long-time limit, where the momentum at every jump is approximately renewed according to $G(\Delta p)$. 
Therefore,  the theory of the exponential well describes the propagator for the HRW model. 

\subsection{Ensemble and time averages of observables }

In this subsection, we consider the ensemble average of an observable, which is defined as  
\begin{eqnarray}
\langle {\mathcal O}(p(t)) \rangle &\equiv& \int_{-p_{\rm trap}}^{p_{\rm trap}} {\mathcal O}(p) \rho(p,t)dp.
\label{ensemble-ave-def}
\end{eqnarray}
We assume that the observable is ${\mathcal O}(p) = C|p|^\beta$ and $\beta>-1$. 
For example, if $\beta=2$ we are considering the kinetic energy of atom. 
Through a change of variables ($p'=t^{\alpha^{-1}} p/c^{\alpha^{-1}}$) and using the scaling function, Eq.~(\ref{sf-exp}),
we have 
\begin{eqnarray}
\langle {\mathcal O}(p(t)) \rangle \sim \int_{- \left(\frac{t}{c}\right)^{\alpha^{-1}} p_{\rm trap}}^{\left(\frac{t}{c}\right)^{\alpha^{-1}} p_{\rm trap}}
 {\mathcal O} \left( \frac{c^{\alpha^{-1}} p'}{t^{\alpha^{-1}}} \right) g_{\rm exp} (p')dp'
\label{ensemble-ave-def-exp}
\end{eqnarray}
for $t\to\infty$. 

When $|p|^\beta$ is integrable with respect to $g_{\rm exp}(p)$, i.e., 
$\int_{-\infty}^\infty g_{\rm exp}(p) |p|^\beta dp<\infty$, $\beta$ satisfies $-1<\beta < \alpha -1$. 
In this case, the asymptotic behavior of the ensemble average becomes 
\begin{equation}
\langle {\mathcal O}(p(t)) \rangle \sim \frac{C c^{\beta \alpha^{-1}}}{t^{\beta\alpha^{-1}}} 
 \int_{-\infty}^\infty |p'|^\beta g_{\rm exp}(p')dp' \quad (t \to \infty).
\label{en-ave-scaling}
\end{equation}
On the other hand, when $|p|^\beta$ is integrable with respect to $I_{\rm exp} (p)$, i.e., 
$\int_{-p_{\rm trap}}^{p_{\rm trap}} I_{\rm exp} (p) {\mathcal O}(p) dv < \infty$,  $\beta$ satisfies $\beta > \alpha-1~(>0)$, 
implying that $|p|^\beta$ is not 
integrable with respect to the scaling function, i.e., $\int_{-\infty}^\infty g_{\rm exp}(p) |p|^\beta dp=\infty$. 
In this case, the asymptotic behavior of the ensemble average becomes 
\begin{equation}
 \langle {\mathcal O}(p(t)) \rangle \sim  t^{\alpha^{-1}-1} \int_{-p_{\rm trap}}^{p_{\rm trap}} I_{\rm exp} (p) {\mathcal O}(p) dv \quad (t \to \infty).
 \label{en-ave-infty-exp}
\end{equation}
Therefore, the asymptotic behavior becomes  
\begin{equation}
\langle {\mathcal O}(p(t)) \rangle \propto t^{-\lambda(\alpha,\beta)}\quad (t \to \infty), 
\end{equation}
and 
the integrability of the observable with respect to the scaling function or infinite invariant density 
determines the power-law exponent $\lambda(\alpha,\beta)$. 
In the case of $\beta = \alpha -1$, the integrals of the observable with respect to both the scaling function and 
infinite invariant density diverge. In this case, the integration in Eq.~(\ref{ensemble-ave-def-exp}) contains a logarithmic 
divergence for $t\to\infty$. Therefore, the leading order for $t\to\infty$ is 
\begin{equation}
\langle {\mathcal O}(p(t)) \rangle \propto t^{\alpha^{-1}-1} \ln t.
\end{equation}

The power-law exponent $\lambda(\alpha,\beta)$ in the exponential model is given by
\begin{equation}
\lambda(\alpha,\beta) = \left\{
\begin{array}{ll}
1 - \alpha^{-1}   & (\beta > \alpha-1)\\
\\
\beta \alpha^{-1} & (\beta < \alpha-1) .
\end{array}
\right.
\label{decay-exp}
\end{equation}
As will be shown later,  the decay process is universal in the sense that $\lambda(\alpha,\beta)$ 
does not depend on the three models that we consider here. Moreover, the fastest decay, which implies the maximum 
of $\lambda(\alpha,\beta)$, is realized at the transition point between integrable and non-integrable with respect to the 
infinite invariant measure, i.e., $\alpha=\beta + 1$. 
In particular, 
the fastest decay of the kinetic energy, i.e., $\beta=2$, can be achieved for $\alpha=3$, 
which suggests that the cooling efficiency, in a sense, is optimized at this point.  As shown in the previous subsection, 
the height of the cooled peak increases with $t^{\alpha^{-1}}$. Moreover, 
the half-width of the cooled peak in the momentum distribution decays with $t^{-\alpha^{-1}}$. 
If we use the half-width of the cooled peak in the momentum distribution to characterize the cooling efficiency, 
the optimized parameter is $\alpha=1$. Therefore, the most efficient cooling parameter depends on the definition of efficiency.

\subsection{Distributional characteristics of time-averaged observables}
Here, we construct a theory of the distribution of time averages in the exponential model. 
The time average of an observable ${\mathcal O}(p)$ is defined by
\begin{equation}
\overline{{\mathcal O}}(t) \equiv \frac{1}{t} \int_0^t {\mathcal O}(p(t'))dt'.
\label{ta-def}
\end{equation}
We obtain the mean and variance for two cases, when the observable
 is integrable with respect to the infinite invariant density and when it is not. In what follows, we consider kinetic energy as a specific example, i.e., ${\mathcal O}(p)=p^2$. 
The integrated value  of an observable ${\mathcal O}(p)$ denoted by ${\mathcal S}(t)$ 
can be represented by
\begin{eqnarray}
{\mathcal S}(t) &=& \int_0^t {\mathcal O}(p(t'))dt'\\
&=& \sum_{i=1}^{N(t)} \Delta {\mathcal S}_i + {\mathcal O}(p_{N(t)+1}) (t-t_{N(t)}),
\end{eqnarray}
where $\Delta {\mathcal S}_i = {\mathcal O}(p_i) \tilde{\tau}_i $, 
$N(t)$ is the number of jumps until time $t$, $p_i$ is the momentum during $[t_{i-1}, t_{i})$, and 
$t_i=\tilde{\tau}_1 + \cdots \tilde{\tau}_i$. 
The integrated value ${\mathcal S}(t)$ is a piecewise linear function of $t$ \cite{Barkai2021,*barkai2022gas} 
because ${\mathcal O}(p(t))$ is a piecewise constant function,
where $p_i$ and $\tilde{\tau}_i$ are coupled stochastically. 
The joint PDF of $\Delta {\mathcal S}_i$, $\tilde{\tau}_i$, and $p_i$ denoted by $\phi_{3} (x,\tilde{\tau},p)$ is given by 
\begin{equation}
\phi_{3} (x,\tilde{\tau},p) = \chi (p) R(p) e^{-R(p) \tilde{\tau}} \delta (x- {\mathcal O}(p)\tilde{\tau}).
\end{equation} 
The joint PDF of the integrated value of an elementary step and the waiting time $\tilde{\tau}$ is given by 
\begin{eqnarray}
\phi_{2} (x, \tilde{\tau}) &=& \int_{-p_{\rm trap}}^{p_{\rm trap}} dp \phi_{3} (x,\tilde{\tau},p) \nonumber\\
&=& \frac{1}{2 p_{\rm trap}\sqrt{x\tilde{\tau}}} R(\sqrt{x/\tilde{\tau}}) e^{-R(\sqrt{x/\tilde{\tau}}) \tilde{\tau}} \quad (\sqrt{x/\tilde{\tau}}<p_{\rm trap}). \nonumber
\end{eqnarray} 

Let $Q(x,t)$ be the PDF of $x={\mathcal S}(t)$ when a jump occurs exactly at time $t$; then, we have
\begin{equation}
Q(x,t) =  \int_0^x dx' \int_0^t dt'  \phi_{t} (x', t') Q(x-x', t-t') + Q_0(x,t), 
\end{equation}
where  $Q_0(x,t)=\delta(x)\delta(t)$. 
The PDF of ${\mathcal S}(t)$ at time $t$ is given by
\begin{eqnarray}
P(x,t) &=&  \int_0^x dx' \int_0^t  dt' \Phi_{2} (x', t') Q(x-x', t-t'),  
\end{eqnarray}
where
\begin{equation}
\Phi_2 (x,t) = \int_t^\infty d\tilde{\tau} \int_{-p_{\rm trap}}^{p_{\rm trap}} dp \chi (p) R(p) e^{-R(p) \tilde{\tau}} \delta (x- {\mathcal O}(p)t) .
\end{equation}
The double-Laplace transform with respect to $x$ and $t$ ($u\leftrightarrow x$ and $s\leftrightarrow t$) yields
\begin{equation}
\widehat{P}(u,s) =  \frac{\widehat{\Phi}_{2}(u,s)}{1- \widehat{\phi}_{2}(u,s)}, 
\label{montroll-weiss-like}
\end{equation}
where $\widehat{\phi}_{2}(u,s)$ and $\widehat{\Phi}_{2}(u,s)$ are 
the double-Laplace transforms of $\phi_{2} (x, \tilde{\tau} )$ and $\Phi_{2} (x,t)$, which are given by 
\begin{eqnarray}
\widehat{\phi}_{2}(u,s) &=& \int_0^\infty dx \int_0^\infty d\tau \int_{-p_{\rm trap}}^{p_{\rm trap}} dp  e^{-ux-s\tau} \phi_3(x,\tau,p)\nonumber\\
&=& \int_0^{p_{\rm trap}} \frac{c^{-1}p_{\rm trap}^{-1}p^\alpha}{s+up^2 + c^{-1}p^\alpha}dp
\label{psi_laplace_ta}
\end{eqnarray}
and
\begin{eqnarray}
\widehat{\Phi}_{2}(u,s) &=& \int_0^{p_{\rm trap}} \frac{p_{\rm trap}^{-1}}{s+up^2 + c^{-1}p^\alpha}dp, 
\label{PSI_laplace_ta}
\end{eqnarray}
respectively. Eq.~(\ref{montroll-weiss-like}) is the exact form of the PDF of ${\mathcal S}(t)$ in Laplace space.  Because 
$1-\widehat{\phi}_{2}(0,s)=s\widehat{\Phi}_{2}(0,s)$, normalization is actually satisfied, i.e., $\widehat{P}(0,s)=1/s$. 

The Laplace transform of the first moment of ${\mathcal S}(t)$ can be obtained as
\begin{equation}
-\left. \frac{\partial \widehat{P}(u,s)}{\partial u} \right|_{u=0} = 
-\frac{\widehat{\Phi}_{2}'(0,s)}{1- \widehat{\phi}_{2}(0,s)} - \frac{\widehat{\phi}_{2}'(0,s)}{s[1- \widehat{\phi}_{2}(0,s)]}. 
\label{laplace-1st-moment-exp}
\end{equation}
For $\alpha<3$, $\widehat{\phi}_{2}'(0,0)$ is finite, whereas it diverges for $\alpha\geq 3$. Therefore, $\alpha=3$ is a transition point 
at which the asymptotic behavior of $\langle {\mathcal S}(t) \rangle$ exhibits a different form.  
The asymptotic behavior of $1- \widehat{\phi}_{2}(0,s)$ for $s\to 0$ is given by 
\begin{eqnarray}
1- \widehat{\phi}_{2}(0,s) &=& s \int_0^{p_{\rm trap}} \frac{p_{\rm trap}^{-1}}{s + cp^\alpha}dp 
\sim A_\alpha s^{1/\alpha},
\end{eqnarray}
where $A_\alpha$ is given by
\begin{equation}
A_\alpha = \frac{c^{1/\alpha}p_{\rm trap}^{-1}\pi}{\alpha \sin (\pi/\alpha)}.
\end{equation}
For $\alpha<3$, the leading order of Eq.~(\ref{laplace-1st-moment-exp}) is 
\begin{equation}
-\left. \frac{\partial \widehat{P}(k,s)}{\partial u} \right|_{u=0} \sim -
 \frac{\widehat{\phi}_{2}'(0,0)}{A_\alpha s^{1+\frac{1}{\alpha}}}, 
 \label{pd-P}
\end{equation}
where the first term in Eq.~(\ref{laplace-1st-moment-exp}) is ignored because $\widehat{\Phi}_{2}'(0,s)\propto s^{3/\alpha -2}$.
Therefore, the asymptotic behavior of $\langle {\mathcal S}(t) \rangle$ becomes 
\begin{equation}
\langle {\mathcal S}(t) \rangle \sim
 \frac{-\widehat{\phi}_{2}'(0,0)}{A_\alpha \Gamma(1+1/\alpha) } t^{\frac{1}{\alpha}}
 \label{1st-moment-X-a<3}
\end{equation}
for $t\to\infty$, where $-\widehat{\phi}_{2}'(0,0)=cp_{\rm trap}^{2-\alpha}/(3-\alpha)$.

For $\alpha \geq 3$, on the other hand, the asymptotic behavior of $\langle {\mathcal S}(t) \rangle$ becomes 
different from Eq.~(\ref{1st-moment-X-a<3}). 
For $\alpha >3$, the asymptotic behaviors of $-\widehat{\phi}_{2}'(0,s)$ and $-\widehat{\Phi}_{2}'(0,s)$ 
 for $s\to 0$ become 
\begin{eqnarray}
-\widehat{\phi}_{2}'(0,s) &=& \int_0^{p_{\rm trap}} \frac{cp_{\rm trap}^{-1}p^{2+\alpha}}{(s + cp^\alpha)^2}dp 
\sim b_\alpha s^{3/\alpha -1}
\end{eqnarray}
and 
\begin{eqnarray}
-\widehat{\Phi}_{2}'(0,s) &=& \int_0^{p_{\rm trap}} \frac{p_{\rm trap}^{-1}p^{2}}{(s + cp^\alpha)^2}dp 
\sim B_\alpha s^{3/\alpha -2},
\end{eqnarray}
where $b_\alpha$ and $B_\alpha$ are given by
\begin{equation}
b_\alpha = \frac{3c^{3/\alpha }\pi p_{\rm trap}^{-1}}{\alpha^2 \sin (3\pi/\alpha)}
\end{equation}
and
\begin{equation}
B_\alpha = \frac{(\alpha -3) \pi p_{\rm trap}^{-1}c^{3/\alpha}}{\alpha^2 \sin (3\pi/\alpha)},
\end{equation}
respectively.
Note that there is a logarithmic correction in the asymptotic behavior of $\langle {\mathcal S}(t) \rangle$ when $\alpha=3$. 
Therefore, the asymptotic behavior of $\langle {\mathcal S}(t) \rangle$ becomes 
\begin{eqnarray}
\langle {\mathcal S}(t) \rangle &\sim&
 \frac{b_\alpha+B_\alpha}{A_\alpha \Gamma(2-2/\alpha) } t^{1-\frac{2}{\alpha}}\nonumber\\
 &=& \frac{c^{2/\alpha}\sin(\pi/\alpha)}{ \Gamma(2-2/\alpha) \sin (3\pi/\alpha) } t^{1-\frac{2}{\alpha}}
 \label{1st-moment-X-a>3}
\end{eqnarray}
for $t\to\infty$. 

The Laplace transform of the second moment of ${\mathcal S}(t)$ can be obtained as
\begin{eqnarray}
\left. \frac{\partial^2 \widehat{P} (u,s)}{\partial u^2} \right|_{u=0} = 
\frac{\widehat{\Phi}_{2}''(0,s)}{1- \widehat{\phi}_{2}(0,s)} + \frac{2\widehat{\Phi}_{2}'(0,s)\widehat{\phi}_{2}'(0,s)}{[1- \widehat{\phi}_{2}(0,s)]^2} \nonumber\\
+\frac{\widehat{\Phi}_{2}(0,s)\widehat{\phi}_{2}''(0,s)}{[1- \widehat{\phi}_{2}(0,s)]^2} + 
 \frac{2\widehat{\Phi}_{2}(0,s)\widehat{\phi}_{2}'(0,s)^2}{[1- \widehat{\phi}_{2}(0,s)]^3}. 
\label{laplace-2nd-moment-exp}
\end{eqnarray}
For $\alpha<3$, the last term represents the leading term. Therefore, we have
\begin{eqnarray}
\left. \frac{\partial^2 \widehat{P}(k,s)}{\partial u^2} \right|_{u=0} \sim
 \frac{2\widehat{\phi}_{2}'(0,0)^2}{s[1- \widehat{\phi}_{2}(0,s)]^2}\sim 
  \frac{2\widehat{\phi}_{2}'(0,0)^2}{A_\alpha^2 s^{1+2/\alpha} }
\end{eqnarray}
for $s\to 0$. It follows that the asymptotic behavior of $\langle {\mathcal S}(t)^2 \rangle$ becomes 
\begin{equation}
\langle {\mathcal S}(t)^2 \rangle \sim
 \frac{2\widehat{\phi}_{2}'(0,0)^2}{A_\alpha^2 \Gamma(1+2/\alpha) } t^{\frac{2}{\alpha}}
 \label{2nd-moment-X-a<3}
\end{equation}
for $t\to\infty$. Because the ergodicity breaking (EB) parameter is given by 
\begin{equation}
{\rm EB}\equiv 
\frac{\langle \overline{\mathcal O}(t)^2 \rangle - \langle \overline{\mathcal O}(t) \rangle^2}{\langle \overline{\mathcal O}(t)\rangle^2} 
= \frac{\langle {\mathcal S}(t)^2 \rangle - \langle {\mathcal S}(t) \rangle^2}{\langle {\mathcal S}(t) \rangle^2}, 
\end{equation}
we have the EB parameter for the kinetic energy:
\begin{equation}
{\rm EB}\to 
 \frac{2 \Gamma(1+1/\alpha)^2}{\Gamma (1+2/\alpha)} -1 
 \label{eb-ML}
\end{equation}
for $t\to \infty$. This is a consequence of the Darling-Kac theorem \cite{Darling1957}. Thus, this is 
a universal result that does not depend on the subrecoil laser cooling model considered here. 

On the other hand, for $\alpha\geq 3$, all the terms in Eq.~(\ref{laplace-2nd-moment-exp}) contribute  to
the asymptotic behavior of $\langle {\mathcal S}(t)^2 \rangle$. 
For $\alpha >3$, the asymptotic behaviors of $\widehat{\Phi}_{2}''(0,s)$ and $\widehat{\phi}_{2}''(0,s)$ for $s\to 0$ become 
\begin{eqnarray}
\widehat{\phi}_{2}''(0,s) &=& \int_0^{p_{\rm trap}} \frac{2c^{-1}p_{\rm trap}^{-1}p^{4+\alpha}}{(s + c^{-1}p^\alpha)^3}dp 
\sim c_\alpha s^{5/\alpha -2}
\end{eqnarray}
and 
\begin{eqnarray}
\widehat{\Phi}_{2}''(0,s) &=& \int_0^{p_{\rm trap}} \frac{2p_{\rm trap}^{-1}p^{4}}{(s + c^{-1}p^\alpha)^3}dp 
\sim C_\alpha s^{5/\alpha -3},
\end{eqnarray}
where $c_\alpha$ and $C_\alpha$ are given by \begin{equation}
c_\alpha = \frac{5(-5+\alpha) \pi p_{\rm trap}^{-1} c^{5/\alpha}}{\alpha^3 \sin (5\pi /\alpha)}
\end{equation}
and 
\begin{equation}
C_\alpha = \frac{(-5+\alpha)(-5+2\alpha) \pi p_{\rm trap}^{-1} c^{5/\alpha}}{\alpha^3 \sin (5\pi /\alpha)},
\end{equation}
respectively. 
It follows that 
\begin{eqnarray}
\left. \frac{\partial^2 \widehat{P}(u,s)}{\partial u^2} \right|_{u=0} \sim
\left(\frac{c_\alpha + C_\alpha}{A_\alpha} + \frac{2B_\alpha b_\alpha}{A_\alpha^2} 
+ \frac{2b_\alpha^2}{A_\alpha^2} \right)s^{4/\alpha-3}\nonumber
\end{eqnarray}
for $s\to 0$. Therefore, in the long-time limit, 
\begin{equation}
\langle {\mathcal S}(t)^2 \rangle \sim \left(\frac{c_\alpha + C_\alpha}{A_\alpha} + \frac{2B_\alpha b_\alpha}{A_\alpha^2} 
+ \frac{2b_\alpha^2}{A_\alpha^2} \right) 
 \frac{t^{2(1-\frac{2}{\alpha})}}{\Gamma(3-4/\alpha) } ,
 \label{2nd-moment-X-a>3}
\end{equation}
 and the EB parameter becomes 
\begin{equation}
{\rm EB}\to \frac{2 \Gamma (2-2/\alpha)^2}{\alpha\Gamma (3-4/\alpha)} 
\left[ \frac{(-5+\alpha) \sin^2 (3\pi/\alpha)}{\sin (5\pi/\alpha) \sin (\pi/\alpha)} +3\right]
  -1 
  \label{EB-p2-a>3}
\end{equation}
for $t\to \infty$. Contrary to the universality in the case of $\alpha<3$, as will be shown later, 
this result is different from that in the deterministic model.

\begin{figure}
\includegraphics[width=.95\linewidth, angle=0]{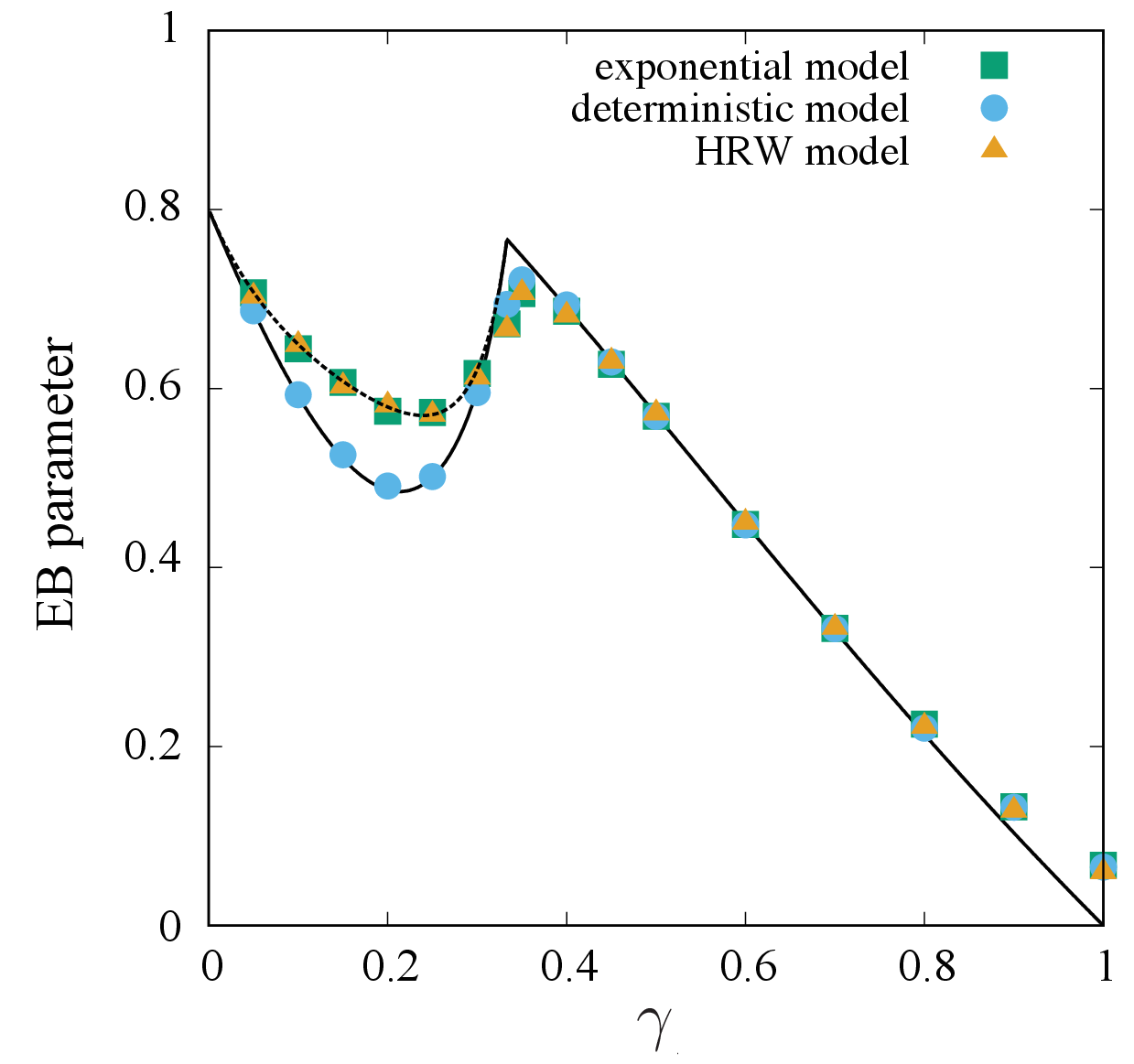}
\caption{EB parameter as a function of $\gamma$ ($=1/\alpha$) for the kinetic energy, i.e., ${\mathcal O}(p)=p^2$. 
Symbols are the results of numerical simulations for the HRW, deterministic, and exponential models. 
The solid line represents ${\rm A}(\gamma)$ and ${\rm ML}(\gamma)$ for $\gamma<1/3$ and $\gamma>1/3$, respectively.
The dashed line represents  Eq.~(\ref{EB-p2-a>3}). The solid line represents Eq.~(\ref{eb-ML}) and (\ref{eb-p2-det}).
 }
\label{eb-gamma}
\end{figure}

\section{Stochastic model with a deterministic coupling}
Here, we consider a stochastic model with a deterministic coupling, i.e., the deterministic model. 
This model is obtained by replacing the conditional PDF of the waiting time given the momentum 
by its mean. 
In this sense, this model is a mean-field-like model of the exponential model.
In the deterministic model, the conditional PDF $q(\tilde{\tau}|p)$ of $\tilde{\tau}$ given $p$ becomes deterministic:  
\begin{equation}
q(\tilde{\tau}|p) = \delta(\tilde{\tau} - R(p)^{-1}). 
\end{equation}
Using Eq.~(\ref{joint-pdf exp}) and integrating over momentum $p$ yields that the PDF of the waiting time follows a power law: 
\begin{equation}
\psi (\tilde{\tau}) = \gamma p_{\rm trap}^{-1} c^\gamma \tilde{\tau}^{-1 -\gamma} \quad (\tilde{\tau}\geq c p_{\rm trap}^{-\gamma^{-1}}).
\label{waiting-time-pdf-det}
\end{equation} 

\subsection{Scaling function and infinite invariant density}
The deterministic model is described by the SMP. Using Eq.~(\ref{MW-SMP}), 
we have
\begin{equation}
\hat{\rho} (p,s) = \frac{\chi(p)}{s} \frac{1 - e^{-sR(p)}}{1-\hat{\psi}(s)}.
\end{equation}
Because $\psi(\tilde{\tau})$ follows a power law, i.e., Eq.~(\ref{waiting-time-pdf-det}), the asymptotic form of the 
the Laplace transform $\hat{\psi}(s)$ for $s\to 0$ is given by 
\begin{equation}
\hat{\psi}(s) = 1 - a s^\gamma + o(s^\gamma), 
\end{equation}
where $a= \Gamma(1-\gamma) p_{\rm trap}^{-1}c^\gamma$. 
In the long-time limit, the propagator is expressed as 
\begin{equation}
\rho(p,t)  \sim 
\begin{cases}
\dfrac{\sin (\pi \gamma) }{2\pi \gamma    } \left(\dfrac{t}{c}\right)^{\gamma } \quad &(|p| \leq p_c(t))\\
\\
\dfrac{\sin (\pi \gamma)}{2 \pi\gamma  } \dfrac{t^{\gamma} - (t- t_{c}(p))^\gamma }{c^\gamma} &(|p|>p_c(t)),
\end{cases}
\label{propagator_asympt1}
\end{equation}
 where $p_{c}(t)=(t/c)^{-\gamma}$ and $t_c(p)= c|p|^{-\gamma^{-1}}$. 
We note that $\rho (p,t)$ is discontinuous at $|p|=p_{c}(t)$, in contrast to the HRW model. 
Importantly, the asymptotic behavior of the propagator, as expressed by Eq.~(\ref{propagator_asympt1}), 
does not depend on the details of the uniform approximation; i.e., $\rho(p,t)$ is independent of $p_{\rm trap}$.
For any small $\varepsilon>0$, there exists $t$ such that $p_c(t) < \varepsilon$ because $p_c(t) \to 0$ for $t\to\infty$. 
Therefore, for any small $\varepsilon>0$, the probability of $|p|>\varepsilon$ becomes zero for $t\to\infty$. 
More precisely, for $t\gg t_c(\varepsilon)$, the probability is given by
\begin{equation}
\Pr (|p|>\varepsilon) \sim \frac{ \sin (\pi \gamma)}{1-\gamma}(1-\varepsilon^{1-\gamma})t^{\gamma -1}.
\end{equation}
Therefore, the temperature of the system almost certainly approaches zero in the long-time limit. 

By changing the variables ($p'=t^{\gamma} p/c^\gamma$), we obtain the rescaled propagator $\rho_{\rm res} (p',t)$. In the long-time 
limit, the rescaled propagator converges to a time-independent function $g_{\rm det} (p')$ (scaling function):
\begin{equation}
 \rho_{\rm res} (p',t) \equiv \rho (c p'/t^{\gamma},t) \left| \frac{dp}{dp'}\right|
\to g_{\rm det} (p')  ,
\label{rescaling}
\end{equation}
where the scaling function is given by
\begin{equation}
 g_{\rm det} (p') \equiv \left\{
\begin{array}{ll}
 \dfrac{  \sin (\pi \gamma )}{2\pi c^{\gamma-1} \gamma   } ~&(|p'|<1)\\
 \\
 \dfrac{ \sin (\pi \gamma ) \{ 1 - (1-|p'|^{-\gamma^{-1}})^\gamma\}}{2\pi c^{\gamma-1} \gamma  } 
 &(|p'| \geq 1) .
 \end{array}
 \right.
 \label{master-curve}
\end{equation}
This scaling function describes the details of the propagator near $p=0$. Furthermore, an infinite invariant density
 is obtained as a formal steady state: 
\begin{equation}
 I_\infty(p) \equiv  \lim_{t\to \infty} t^{1-\gamma} \rho(p,t) 
 = \frac{ \sin (\pi \gamma ) \left\vert p\right\vert ^{-\gamma^{-1}}}{2 \pi c^\gamma  }
\label{inf-d}
\end{equation}
for $|p|<p_{\rm trap}$.  In the long-time limit, the propagator can be almost described by the infinite invariant density, 
 whereas the former is normalized and the latter is not. The 
infinite invariant density $I_\infty(p)$ is the same as the formal steady state obtained using Eq.~(\ref{steady-state}). 
However,  the propagator described by Eq.~(\ref{propagator_asympt1}) is not a solution of the master equation, 
Eq.~(\ref{Master1}).

\begin{figure}
\includegraphics[width=.95\linewidth, angle=0]{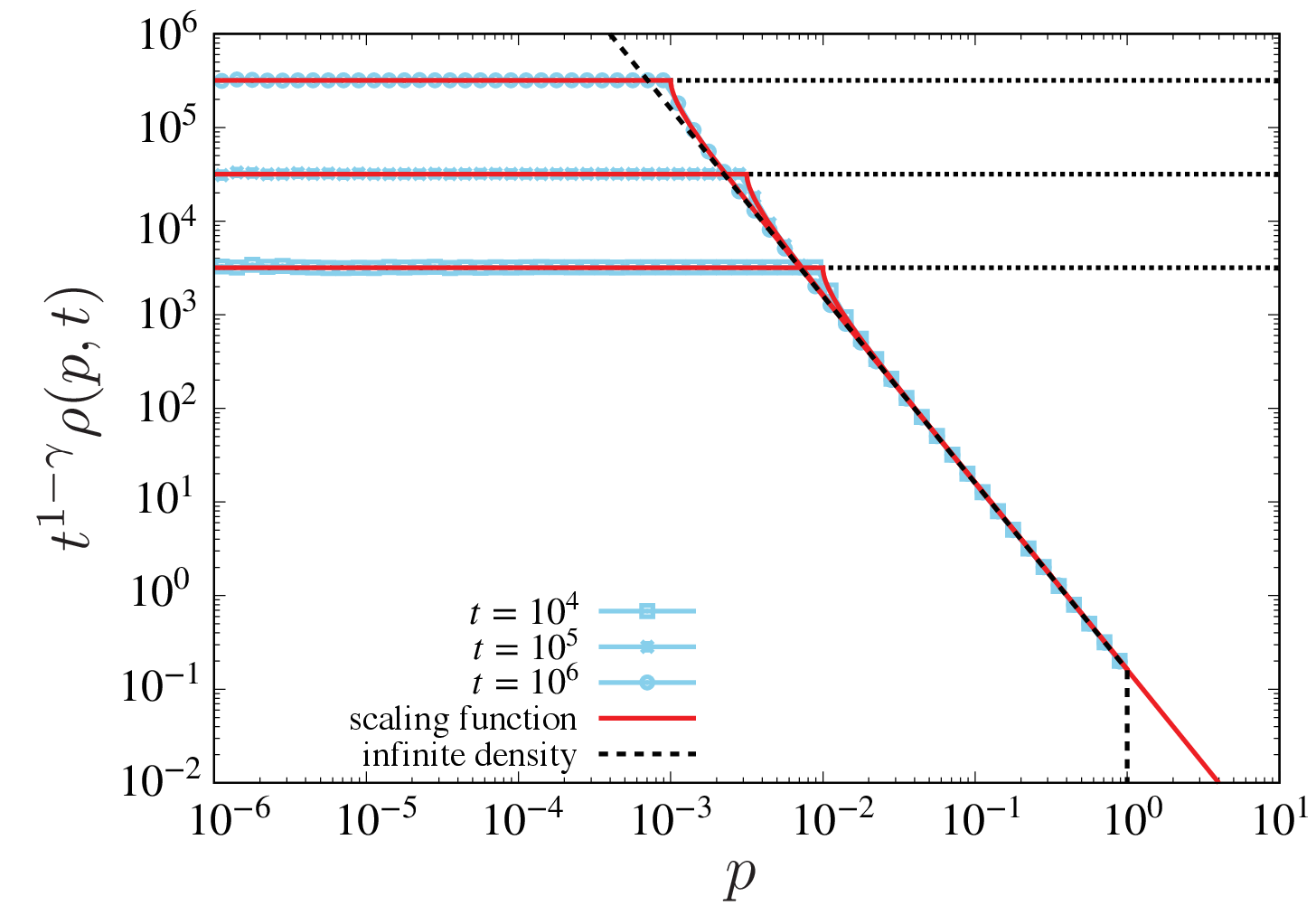}
\caption{Time evolution of the propagator multiplied by $t^{\gamma-1}$ in the deterministic model 
for different times ($\alpha=\gamma^{-1} =2, c=1$, and $p_{\rm trap}=1$). 
Symbols with lines represent the results of numerical simulations of the deterministic model. 
The dashed lines represent the infinite invariant density $I_\infty(p)$ given by Eq.~(\ref{inf-d}). 
The solid lines represent rescaled scaling functions, 
$t g_{\rm det} (t^\gamma p)$. The dotted lines represent $t g_{\rm det} (0)$ for different values of $t$. 
The initial position is chosen uniformly on $[-1,1]$.  
 }
\label{propagator}
\end{figure}

\begin{figure}
\includegraphics[width=.9\linewidth, angle=0]{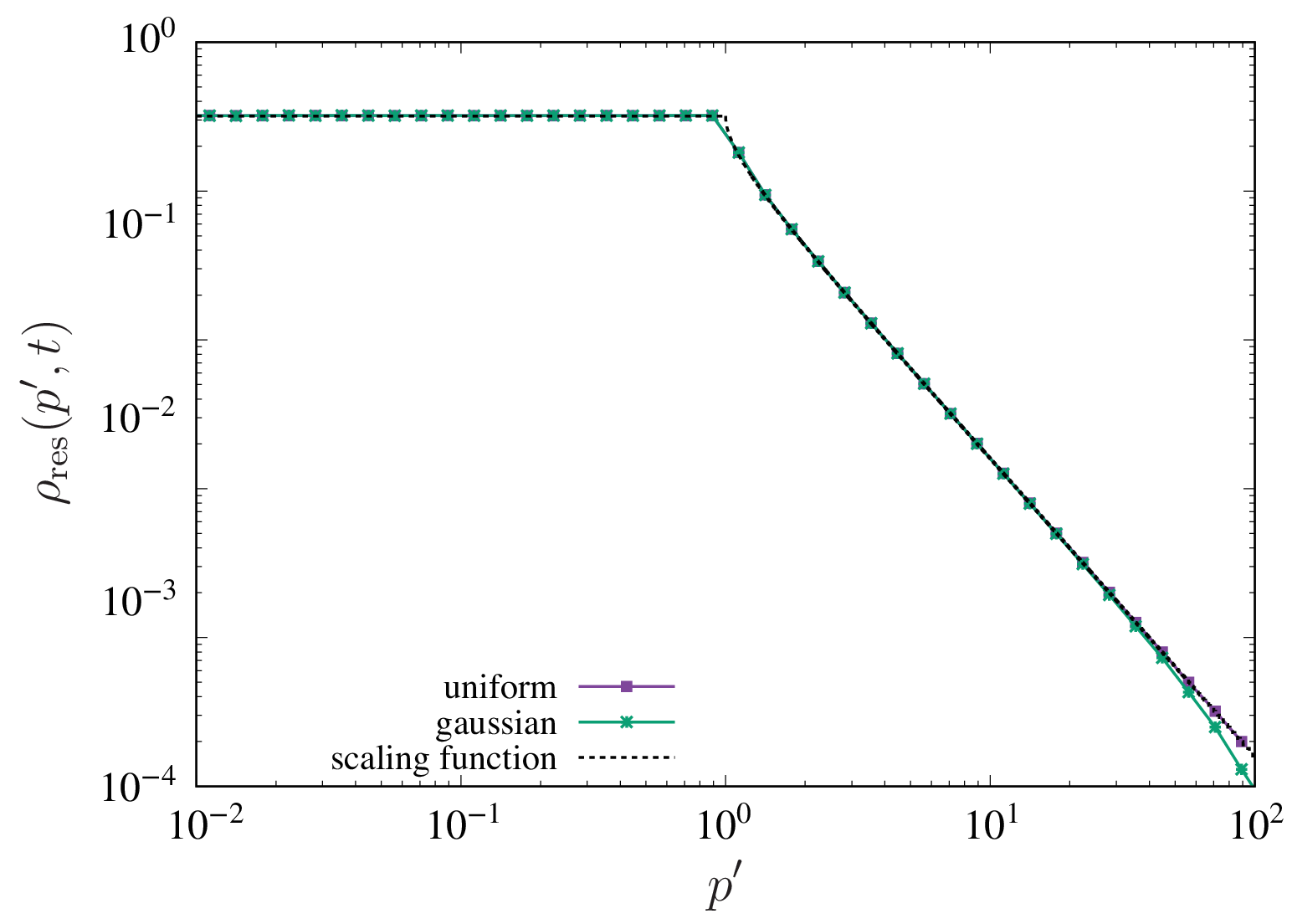}
\caption{Rescaled propagators  for different distributions $\chi(p)$ ($\alpha=\gamma^{-1} =2$, $c=1$,  and $p_{\rm trap}=1$), where 
we consider the uniform distribution $\chi(p)=1/2$ on $p\in [-1,1]$ and the Gaussian distribution $\chi (p) = 
\exp(-p^2/2)/\sqrt{2\pi}$. 
Symbols with lines are the results of the numerical simulations of the deterministic model with $t=10^4$. 
The solid line represents the scaling function given by Eq.~(\ref{master-curve}). 
The initial position is chosen uniformly on $[-1,1]$. 
Note that the results for different $\chi(p)$ are indistinguishable.}
\label{propagator-rescale}
\end{figure}

Figure~\ref{propagator} shows the scaled propagator of the deterministic model.
In the numerical simulations, we generated $10^8$ trajectories to obtain the propagator. 
There are two forms of the propagator. For $|p|<p_c(t)$, the propagator increases with 
time $t$. For $|p|>p_c(t)$, the asymptotic form of the propagator follows the infinite invariant density $t^{\gamma-1}I_\infty(p)$. 
Because the constant $t^{\gamma -1}$ approaches zero in the long-time limit, the propagator outside $p_c(t)$ becomes zero. 
A cusp exists at $p=t_c(t)$, in contrast to the HRW and the exponential model, where no cusp exists in the propagator. 
Figure~\ref{propagator-rescale} shows numerical simulations of 
the rescaled propagators in the deterministic case for different $\chi(p)$, i.e., for uniform and Gaussian distributions. The propagators 
are compared with the scaling function $g_{\rm det} (p')$ without fitting parameters, where we generate $10^8$ trajectories 
to obtain the rescaled propagator. 
 Therefore, the scaling function 
describes the details of the propagator near $p=0$ and is universal in the sense that it does not depend on $\chi(p)$.

\subsection{Ensemble and time averages of observables}

Here, we consider the ensemble averages of observables and 
show that the scaling function and infinite invariant density play an important role. 
In this subsection, we set $p_{\rm trap}=1$ for simplicity.
The ensemble average of an observable ${\mathcal O}(p)$ is given by Eq.~(\ref{ensemble-ave-def}), 
which can be represented using the scaling function and infinite invariant density. To verify, we divide the integral range 
as 
\begin{widetext}
\begin{equation}
\langle {\mathcal O}(p(t)) \rangle 
= \int_{-p_c(t)}^{p_c(t)} \rho(p,t) {\mathcal O}(p) dp + \int_{|p|>p_c(t)} \rho (p,t) {\mathcal O}(p)dp.
\label{ensemble-ave}
\end{equation}
In the long-time limit, using the scaling function and infinite invariant density, we have 
\begin{equation}
\langle {\mathcal O}(p(t)) \rangle 
\cong \int_{-1}^{1} g_{\rm det} (p') {\mathcal O}(cp'/t^{\gamma}) dp' + t^{\gamma-1} \int_{|p|>p_c(t)} I_\infty (p) {\mathcal O}(p) dp,
\label{ensemble-ave2}
\end{equation}
where we applied a change of variables in the first term and used Eqs.~(\ref{propagator_asympt1}), (\ref{master-curve}), and 
(\ref{inf-d}). 
\end{widetext}

Here, we assume that ${\mathcal O}(p) \sim C|p|^\beta$ for $p\to 0$ and that it is bounded for $p\ne 0$. In particular,  
the energy and the absolute value of the momentum correspond to observables with 
$\beta=2$ and $\beta=1$, respectively. 
When $|p|^\beta$ is integrable with respect to $g_{\rm det}(p)$, i.e., 
$\int_{-\infty}^\infty g_{\rm det}(p) |p|^\beta dp<\infty$, 
$\gamma^{-1}$ satisfies the following inequality: $-1<\beta < \gamma^{-1}-1$. 
In this case, the asymptotic behavior of the ensemble average becomes 
\begin{equation}
\langle {\mathcal O}(p(t)) \rangle \sim \frac{C c^{\beta -\gamma+1} \sin(\pi\gamma)}{\pi  \gamma(\beta+1)} 
 t^{-\beta \gamma} \quad (t \to \infty), 
\label{en-ave-scaling}
\end{equation}
where we used Eq.~(\ref{master-curve}):
\begin{equation}
\int_{-1}^{1} g_{\rm det}(p') {\mathcal O}(cp'/t^{\gamma})dp' \sim C c^\beta  \int_{-1}^1 g_{\rm det}(p') |p'|^{\beta} 
dp' t^{-\beta \gamma}
\end{equation}
for $t \to \infty$.
Note that the second term in Eq.~(\ref{ensemble-ave2}) can be ignored in the asymptotic behavior because 
$-\beta \gamma>\gamma -1$. 
On the other hand, 
when ${\mathcal O}(p)$ is integrable with respect to $I_\infty (p)$, i.e., $\int_{-1}^1 I_\infty (p) {\mathcal O}(p) dp < \infty$,  
where  $\beta$ must satisfy $\beta > \gamma^{-1}-1~ (>0)$, the asymptotic behavior of the ensemble average becomes 
\begin{equation}
 \langle {\mathcal O}(p(t)) \rangle \sim  t^{\gamma-1} \int_{-1}^1 I_\infty (p) {\mathcal O}(p) dp \quad (t \to \infty).
 \label{en-ave-infty}
\end{equation} 
Therefore, the asymptotic behavior of the ensemble average becomes proportional to $ t^{-\lambda(\alpha,\beta)}$, and 
the integrability of the observable with respect to the scaling function or infinite invariant density 
determines the power-law exponent $\lambda(\alpha,\beta)$. Note that the exponent $\gamma$ is defined as 
$\gamma=1/\alpha$. Therefore, the power-law exponent in 
decay processes of the ensemble- and time-averaged observable is universal. 

In the case of $\beta = \gamma^{-1}-1$, the integrals of the observables with respect to both the scaling function and 
infinite invariant density diverge.
In this case, Eq.~(\ref{ensemble-ave2}) should be expressed as 
\begin{widetext}
\begin{equation}
\langle {\mathcal O}(p(t)) \rangle 
= \int_{-1}^{1} g_{\rm det} (p') {\mathcal O}(cp'/t^{\gamma}) dp' +
\int_{1<|p'|\leq t^\gamma/c} g_{\rm det} (p') {\mathcal O}(cp'/t^{\gamma}) dp' .
\label{ensemble-ave3}
\end{equation}
\end{widetext}
The first term decays as $t^{-\beta \gamma}$ because the integral of the observable ${\mathcal O}(p)$ from -1 to 1 
with respect to the scaling function is finite. 
Because there is a logarithmic correction in the second term, the second term yields the leading order for $t\to\infty$:
\begin{equation}
\langle {\mathcal O}(p(t)) \rangle \sim \frac{C c^{\gamma^{-1}-\gamma-1} \gamma \sin(\pi\gamma)}{\pi  } t^{\gamma-1} \ln t.
\end{equation}

Here, we discuss the decrease of the energy.  When the observable is the energy, i.e., 
${\mathcal O}(p)=p^2$, the asymptotic decay is 
\begin{equation}
\langle p(t)^2 \rangle \sim \frac{t^{-2 \gamma}}{\beta+1}\quad(t\to\infty)
\end{equation} 
or
\begin{equation}
\langle p(t)^2 \rangle \sim t^{\gamma-1} \int_{-1}^1 I_\infty (p) {\mathcal O}(p) dv \quad(t\to\infty)
\end{equation} 
for $\gamma^{-1}>3$ and $\gamma^{-1}<3$, respectively. Thus, the ensemble average of the energy approaches zero in the long-time limit. 
Interestingly, a constraint exists in the power-law exponent
$\lambda(2,\gamma)$; i.e., $\lambda(2,\gamma) \leq 2/3$, where the equality holds at $\gamma^{-1}=\alpha=3$. 
 For general observables, 
the power-law exponent is restricted as 
\begin{equation}
\lambda(\beta,\gamma) < \frac{\beta}{\beta +1}.
\end{equation}
In the case of the absolute value of the momentum, it is bounded as $\lambda(1,\gamma) < 1/2$, which is maximized 
at $\gamma^{-1}=2$. 

\subsection{Distributional characteristics of time-averaged observables}
 Distributional limit theorems for time-averaged observables  in the SMP with continuous state variables 
 were also considered in Ref.~\cite{Akimoto2020},
where the infinite invariant density plays an important role in discriminating classes of observables. 
For the SMP, the integral of ${\mathcal O}(p(t))$ is a piecewise linear function of $t$  and is called a 
  continuous accumulation process \cite{Akimoto2015}. The 
ensemble average of an increment of one segment, i.e., 
\begin{equation}
\left\langle \int_0^{\tilde{\tau}} {\mathcal O}(p(t'))dt'\right\rangle \equiv \int_0^\infty \tilde{\tau} {\mathcal O}
\left( c^\gamma\tilde{\tau}^{-\gamma} \right) \psi (\tilde{\tau})d\tilde{\tau},
\end{equation}
may diverge for some observables. When it is finite, the distribution function of the time-averaged observable 
follows the Mittag--Leffler distribution, which is a well-known distribution in infinite ergodic theory \cite{Aaronson1997,shinkai2006lempel} 
and stochastic processes \cite{Darling1957,kasahara77,Lubelski2008,He2008, Miyaguchi2011, Miyaguchi2013, Akimoto2013a,AkimotoYamamoto2016a,Albers2018, Radice2020, Albers2022}.
On the other hand, when it diverges, other non-Mittag-Leffler limit distributions are known \cite{Akimoto2008, Akimoto2015, Albers2018, Akimoto2020, Barkai2021,*barkai2022gas, Albers2022}.
This condition of integrability of the increment can be represented by 
the integrability of the observable with respect to the infinite invariant density. 

\begin{figure}
\includegraphics[width=.95\linewidth, angle=0]{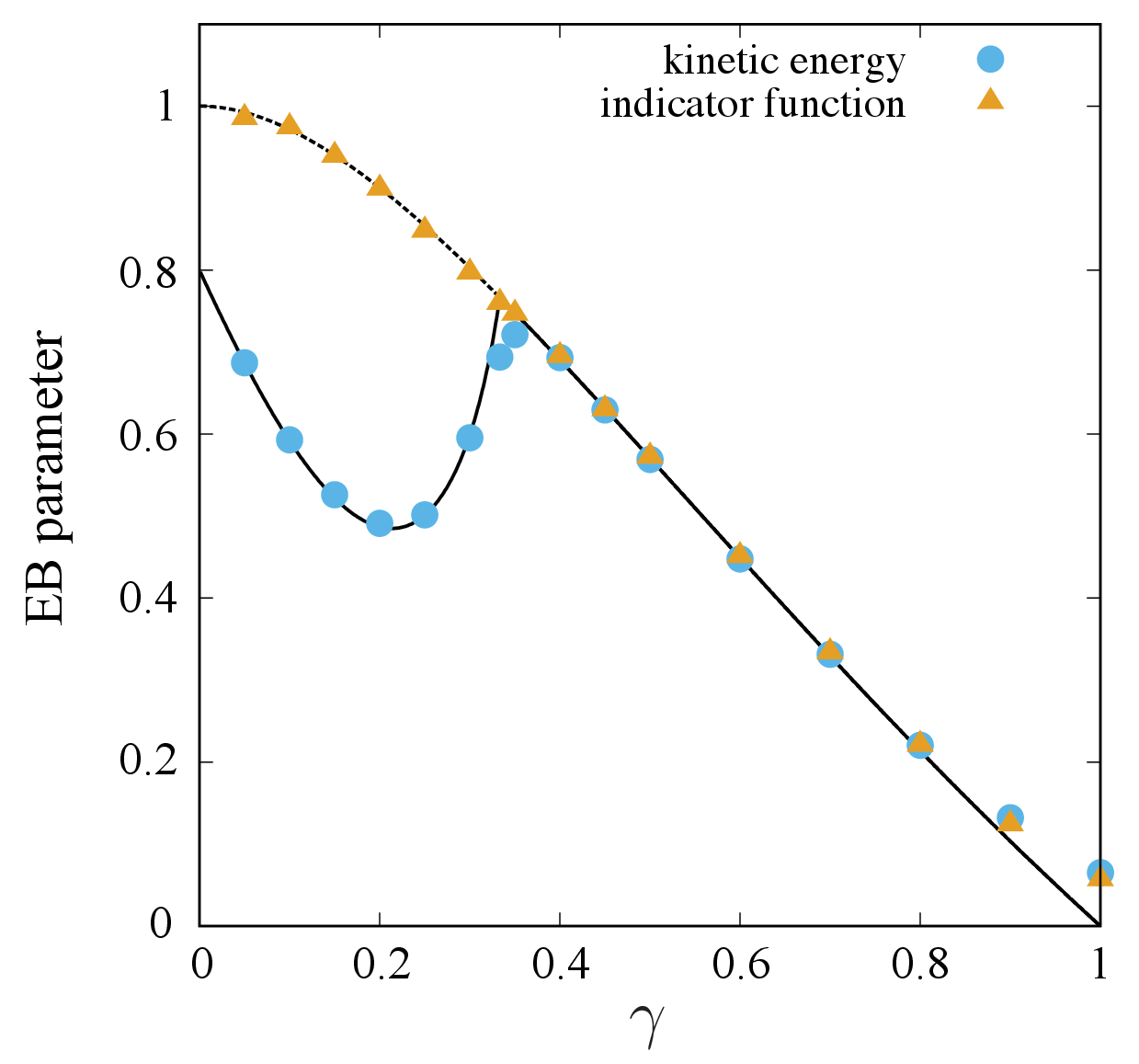}
\caption{EB parameter as a function of $\gamma$ for two observables ${\mathcal O}(p)=p^2$ and ${\mathcal O}(p)=I(|p|>0.5)$, 
where ${\mathcal O}(p)=I(|p|>0.5)=1$ if $|p|>0.5$ and zero otherwise. 
The solid line represents ${\rm A}(\gamma)$ and ${\rm ML}(\gamma)$ for $\gamma<1/3$ and $\gamma>1/3$, respectively.
The dashed line represents  ${\rm ML}(\gamma)$ for $\gamma<1/3$. Note that $I(|p|>0.5)$ is integrable with respect to 
$I_\infty (p)$ for all $\gamma$. 
 }
\label{eb-det}
\end{figure}

Here, we consider energy as a specific example. The distributional limit theorems derived in Ref.~\cite{Akimoto2020} can be 
straightforwardly applied to this case. A derivation of the distributional limit theorems is given in Appendix~A. Here, we 
simply apply our previous results. 
For $\gamma<1/3$, the observable ${\mathcal O}(p)=p^2$ is 
integrable with respect to the infinite invariant density, i.e., $\int_0^1{\mathcal O}(p) I_\infty(p) dp <\infty$, where the ensemble 
average of the increment is finite. Therefore, the distribution of the time average follows the Mittag--Leffler distribution. 
More precisely,  the normalized time averages defined by $\overline{{\mathcal O}}(t)/\langle \overline{\mathcal O}(t)\rangle$ converges 
in distribution: 
\begin{equation}
\frac{\overline{\mathcal O}(t) }{\langle \overline{\mathcal O}(t)\rangle } 
\Rightarrow M_\gamma 
\end{equation}
for $t\to\infty$, where $M_\gamma$ is a random variable, distributed according to the Mittag-Leffler law \cite{Aaronson1997, Miyaguchi2013}.
The ensemble average of the time average decays as 
$\langle \overline{\mathcal O}(t)\rangle \propto t^{\gamma-1}$ for $t\to\infty$ and, in general, 
$\langle \overline{\mathcal O}(t)^n \rangle \propto t^{n(\gamma-1)}$ for $t\to\infty$. Thus, $M_\gamma$ does not depend on time $t$. 
in the long-time limit. 
The mean of $M_\gamma$ is one by definition and the variance 
is given by 
\begin{equation}
{\rm ML}(\gamma) \equiv \frac{2\Gamma(1+\gamma)^2}{\Gamma (1+2\gamma)} -1.
\label{eb-ML2}
\end{equation}
On the other hand, for $\gamma \geq 1/3$, the observable ${\mathcal O}(p)=p^2$ is not integrable with respect to 
the infinite invariant density, and the ensemble average of the increment also diverges. In this case, 
 the normalized time average does not converge in distribution to $M_\gamma$ but rather to another random variable $C_\gamma$ \cite{Akimoto2020}: 
\begin{equation}
\frac{\overline{\mathcal O}(t)}{\langle \overline{\mathcal O}(t)\rangle} 
\Rightarrow C_\gamma 
\end{equation}
for $t\to\infty$. The ensemble average of the time average decays as 
$\langle \overline{\mathcal O}(t)\rangle \propto t^{-2\gamma}$ for $t\to\infty$ and, in general, 
$\langle \overline{\mathcal O}(t)^n \rangle \propto t^{-2n\gamma}$ for $t\to\infty$. The variance of $C_\gamma$ is given by
\begin{equation}
 {\rm A}(\gamma)\equiv \frac{6\gamma \Gamma(2-2\gamma)^2}{\Gamma(3-4\gamma)} 
\left[\frac{ 3 \Gamma (2 - 5\gamma) \Gamma (1-\gamma) }{5\gamma   \Gamma(1 -3\gamma)^2}
+ 1 \right] -1.
\label{eb-p2-det}
\end{equation}
Since the distribution of the normalized time average defined by $\overline{{\mathcal O}}(t)/\langle \overline{\mathcal O}(t)\rangle$ 
converges to $M_\gamma$ or $C_\gamma$ for $\gamma<1/3$ and $\gamma>1/3$, respectively, 
 the EB parameter, which is defined by the relative variance 
of $\overline{\mathcal O}(t)$, i.e., $\langle \overline{\mathcal O}(t)^2 \rangle/\langle \overline{\mathcal O}(t) \rangle^2 -1$. 
is given by ${\rm A}(\gamma)$ and ${\rm ML}(\gamma)$  for $\gamma<1/3$ and $\gamma>1/3$, respectively. 
As shown in Fig.~\ref{eb-det}, the trajectory-to-trajectory fluctuations of $\overline{\mathcal O}(t)$ are suppressed by  
increasing $\gamma$ for $\gamma>1/3$ and vanish for $\gamma\to 1$. On the other hand, they 
show a non-trivial dependence on $\gamma$ for $\gamma<1/3$. 
We note that ${\displaystyle \lim_{\gamma\to1/3}{\rm A}(\gamma) = \lim_{\gamma\to1/3}{\rm ML}(\gamma)}$. 

\begin{table*}
  \begin{tabular}{|p{30mm}|p{30mm}|p{30mm}|p{30mm}|} 
    \hline  & HRW & exponential model & deterministic model \\ \hline
    model & Markov & Markov & non-Markov \\ \hline
    invariant density & $\rho^*(p) \propto |p|^{-\alpha}$ & $\rho^*(p) \propto |p|^{-\alpha}$  & $\rho^*(p) \propto |p|^{-\alpha}$ \\ \hline
    scaling function & same as in the exponential model & Eq. (\ref{sf-exp}) &  Eq.~(\ref{master-curve}) \\ \hline
    decay exponent & same as in the exponential model & Eq. (\ref{decay-exp}) &  Eq. (\ref{decay-exp}) \\ \hline
    EB (integrable) & same as in the exponential model & $\dfrac{2\Gamma(1+\gamma)^2}{\Gamma(1+2\gamma)} -1$ &  $\dfrac{2\Gamma(1+\gamma)^2}{\Gamma(1+2\gamma)} -1$ \\ \hline     
    EB (non-integrable) & same as in the exponential model & Eq. (\ref{EB-p2-a>3}) &  Eq. (\ref{eb-p2-det}) \\ \hline
  \end{tabular}
  \caption{Comparison of the infinite invariant density, the scaling function, the relaxation power-law exponent of the time-and-ensemble averaged energy, and the EB parameter in three stochastic models.}
  \label{sum} 
\end{table*}

\section{Conclusion}
We investigated the accumulation process of the momentum of an atom in three stochastic models of subrecoil laser cooling.
For the HRW and the exponential models, the formal steady state of the master equation
 cannot be normalized when $\alpha\geq 1$. For all the models, the scaled propagator defined by $t^{1-\gamma} \rho (p,t)$ 
 converges to a time-independent function, i.e., an infinite invariant density. 
In the deterministic and exponential model, we derived the exact forms of the scaling function and the 
infinite invariant density. As a result, we found universality and non-universality in all three stochastic models.
In particular, the power-law form of the infinite invariant density 
is universal in the three models, whereas there is a clear difference in the scaling functions of the 
deterministic and exponential models. A summary of the comparisons of the three stochastic models 
is presented in Table~\ref{sum}.

We numerically showed that the propagator obtained using the exponential model 
is in perfect agreement with that in the HRW model for large $t$, which means that 
 the uniform approximation used in the exponential model 
is very useful for obtaining a deeper understanding of the HRW model. 
When we focus on the jumps of the momentum to the trapping region, the jump distribution can be taken as 
approximately uniform in 
the trapping region because the trap size $p_{\rm trap}$ can be arbitrarily small. We note that the uniform distribution for $\chi(p)$ is 
necessary but the value of $p_{\rm trap}$ is not relevant for reproducing the statistical behavior of the HRW model.
This is the reason why the uniform approximation can be applied to the HRW model. 
The relation between the exponential and the HRW models is similar to that between the CTRW and the quenched trap model (QTM)  \cite{bouchaud90}. 
In particular, the waiting times in the exponential model and the CTRW are IID random variables, whereas 
those in the HRW and the QTM are not. Moreover, it is known that the CTRW
is a good approximation of the QTM when the dimension is greater than two or under a bias \cite{Machta1985}.

We showed that the integrability of observables with respect to the infinite invariant density
determines the power-law-decay exponent in the decrease of the ensemble average of the observables 
in the exponential and deterministic models. 
As a result, we found that the power-law exponent has a maximum at the transition point for both models.
Furthermore, we found that 
the integrability of the observable with respect to the infinite invariant density 
plays an important role in characterizing the trajectory-to-trajectory 
fluctuations of the time averages in the three models.  When the observables are integrable, the distribution 
is universal and described by the Mittag-Leffler distribution. 
On the other hand, the distribution differs for the exponential and the deterministic model when the observables are not integrable. 
Using the EB parameter, we numerically showed that the distribution in the HRW model agrees with that in the 
exponential model even when the observable is not integrable.

\section*{Acknowledgement}
T.A. was supported by JSPS Grant-in-Aid for Scientific Research (No.~C JP18K03468). 
The support of Israel Science Foundation's grant 1898/17 is acknowledged (EB).

\appendix

\section{Simulation algorithm}

For all the models, we generate trajectories starting with uniform initial conditions. In the HRW model, the momentum jumps are
generated by random variables following a Gaussian distribution with mean 0 and variance $\sigma^2$ by the Box-Muller's method \cite{box1958note}. 
When momentum becomes $p$ after a momentum jump, the waiting time is a random variable following an exponential distribution with rate $R(p)$. 
In numerical simulations, the waiting time $\tilde{\tau}$ is generated by $\tilde{\tau} = -\ln ({\bm X}) / R(p)$, where ${\bm X}$ is a uniform random variable on $[0,1]$.  
In the HRW model, we consider the reflecting boundary condition at $p=\pm p_{\max}$. In particular, when the momentum becomes $p>p_{\max}$,
  we have  $2p_{\max} - p$. If $p<-p_{\max}$, we have $-2p_{\max} - p$.

For the exponential and deterministic models, 
updates of the momentum are independent of the previous momentum and generated by a uniform random variable on $[-p_{\rm trap}, 
p_{\rm trap}]$. The waiting time in the exponential model is generated in the same way as in the HRW model. 
The waiting time given $p$ in the deterministic model is determined by $\tilde{\tau} =1/R(p)$.

\begin{widetext}
\section{Asymptotic solution to the master equation for the HRW model}
Here, we show that the asymptotic solution of the master equation for the exponential model, i.e., Eq.~(\ref{propagator_asympt-exp2}), 
is also a solution of the master equation for the HRW model. 
Differentiating Eq.~(\ref{propagator_asympt-exp2}) with respect to $t$ gives 
\begin{equation}
\frac{\partial \rho(p,t)}{\partial t}  
\cong - R(p) \rho (p,t) + 
\frac{\sin (\pi \alpha^{-1}) t^{\alpha^{-1}-1 }}{2\pi c^{\alpha^{-1}} \Gamma (\alpha^{-1})} \int_0^1 du 
e^{- R(p) t(1-u)} u^{\alpha^{-1} -1}.
\end{equation}
The first term is the same as that of the master equation of the HRW model, i.e., Eq.~(\ref{Master1}). For $|p|\to 0$, it becomes 
\begin{equation}
\frac{\partial \rho(p,t)}{\partial t}  
\cong - R(p) \rho (p,t) + 
\frac{\sin (\pi \alpha^{-1}) t^{\alpha^{-1}-1 }}{2\pi c^{\alpha^{-1}} \Gamma (1 + \alpha^{-1})} .
\label{propagator-derivative-approx}
\end{equation}
Using Eq.~(\ref{propagator_asympt-exp2}), we approximately calculate the second term of the master equation of the HRW model, i.e., Eq.~(\ref{Master1}).
\begin{equation}
\int_{-p_{\max}}^{p_{\max}} dp^{\prime }\rho \left( p^{\prime },t\right) R(p^{\prime }) \tilde{G}(p|p^{\prime }) \cong 
\frac{\sin (\pi \alpha^{-1}) t^{\alpha^{-1} }}{2\pi c^{\alpha^{-1}} \Gamma (1 + \alpha^{-1})} 
\int_{-\infty}^{\infty} dp^{\prime } G(p-p^{\prime })  \int_0^1 du R(p^{\prime })  e^{-R(p^{\prime })  t(1-u)} u^{\alpha^{-1} -1},
\label{master1-sub}
\end{equation}
where we assumed $|p| , |p^{\prime }| \ll p_{\max}$ and used $\tilde{G}(p|p^{\prime }) \cong G(p-p^{\prime })$. Integrating Eq.~(\ref{master1-sub}) 
by parts, we have 
\begin{equation}
 \int_0^1 du R(p^{\prime })  e^{-R(p^{\prime })  t(1-u)} u^{\alpha^{-1} -1} \cong t^{-1} - (\alpha^{-1} -1) \int_0^1 du e^{-R(p^{\prime })  t(1-u)} u^{\alpha^{-1} -2}
 \sim t^{-1}.
\end{equation}
Thus, Eq.~(\ref{master1-sub}) becomes  
\begin{equation}
\int_{-p_{\max}}^{p_{\max}} dp^{\prime }\rho \left( p^{\prime },t\right) R(p^{\prime }) \tilde{G}(p|p^{\prime }) \cong 
\frac{\sin (\pi \alpha^{-1}) t^{\alpha^{-1}-1 }}{2\pi c^{\alpha^{-1}} \Gamma (1 + \alpha^{-1})} ,
\end{equation}
which is the same as the second term of Eq.~(\ref{propagator-derivative-approx}). Here, we confirmed that Eq.~(\ref{propagator_asympt-exp2}) is a solution 
to the master equation of the HRW model under the assumption of $|p| \to 0$. 
For the HRW model, momentum converges to $p=0$ almost surely in the long-time limit.
Therefore, Eq.~(\ref{propagator_asympt-exp2}) is a solution to the master equation of the HRW model in the long-time limit. 
\end{widetext}

\section{Derivation of the $n$th moment of ${\mathcal S}(t)$}

Here, we derive the $n$th moments of ${\mathcal S}(t)$ for $\alpha<3$ in the exponential model. 
For $\alpha<3$, $\widehat{\phi}_{2}'(0,0) <\infty$. The leading term of the Laplace transform 
of the  $n$th moment is 
\begin{eqnarray}
\left. \frac{\partial^n \widehat{P} (u,s)}{\partial u^2} \right|_{u=0} \sim
 \frac{(-1)^n n!\widehat{\phi}_{2}'(0,s)^n}{[1- \widehat{\phi}_{2}(0,s)]^n}
\label{laplace-nth-moment-exp}
\end{eqnarray}
for $s\to 0$. It follows that the asymptotic behavior of $\langle {\mathcal S}(t)^n \rangle$ becomes 
\begin{equation}
\langle {\mathcal S}(t)^n \rangle \sim
 \frac{n!\{-\widehat{\phi}_{2}'(0,0)\}^n}{A_\alpha^n \Gamma(1+n/\alpha) } t^{\frac{n}{\alpha}}
 \label{nth-moment-X-a<3}
\end{equation}
for $t\to\infty$. In the long-time limit,  the $n$th moment of ${\mathcal S}(t)/\langle {\mathcal S}(t)\rangle$
converges to $n!\Gamma(1+1/\alpha)^n/\Gamma(1+n/\alpha)$ for all $n>0$. Therefore, the random variable defined by 
$\overline{\mathcal S} \equiv {\mathcal S}(t)/\langle {\mathcal S}(t)\rangle$ does not depend on time $t$ 
in the long-time limit and follows the Mittag--Leffler distribution with exponent $1/\alpha$, where 
the Laplace transform of  the random variable $M_\alpha$ following the Mittag--Leffler distribution with exponent $1/\alpha$ is given by 
\begin{equation}
\langle e^{-s  M_\alpha} \rangle = \sum_{k=0}^\infty \frac{\Gamma(1+1/\alpha)^n}{\Gamma(1+n/\alpha)} (-s)^n.
\end{equation} 
In real space, the PDF $f_\alpha(x)$ of $M_\alpha$ becomes
\begin{equation}
f_\alpha (x) = - \frac{1}{\pi \alpha} \sum_{k=1}^\infty \frac{\Gamma (k \alpha + 1)}{k!} x^{k-1} \sin (\pi k \alpha).
\end{equation}


\begin{thebibliography}{58}%
\makeatletter
\providecommand \@ifxundefined [1]{%
 \@ifx{#1\undefined}
}%
\providecommand \@ifnum [1]{%
 \ifnum #1\expandafter \@firstoftwo
 \else \expandafter \@secondoftwo
 \fi
}%
\providecommand \@ifx [1]{%
 \ifx #1\expandafter \@firstoftwo
 \else \expandafter \@secondoftwo
 \fi
}%
%
%
\providecommand \bibnamefont  [1]{#1}%
\providecommand \bibfnamefont [1]{#1}%
\providecommand \citenamefont [1]{#1}%
\providecommand \href@noop [0]{\@secondoftwo}%
\providecommand \href [0]{\begingroup \@sanitize@url \@href}%
\providecommand \@href[1]{\@@startlink{#1}\@@href}%
\providecommand \@@href[1]{\endgroup#1\@@endlink}%
\providecommand \@sanitize@url [0]{\catcode `\\12\catcode `\$12\catcode
  `\&12\catcode `\#12\catcode `\^12\catcode `\_12\catcode `\%12\relax}%
\providecommand \@@startlink[1]{}%
\providecommand \@@endlink[0]{}%
\providecommand \url  [0]{\begingroup\@sanitize@url \@url }%
\providecommand \@url [1]{\endgroup\@href {#1}{\urlprefix }}%
\providecommand \urlprefix  [0]{URL }%
%
\providecommand \doibase [0]{http://dx.doi.org/}%
\providecommand \selectlanguage [0]{\@gobble}%
\providecommand \bibinfo  [0]{\@secondoftwo}%
\providecommand \bibfield  [0]{\@secondoftwo}%
%
\providecommand \BibitemOpen [0]{}%
%
%
%
\providecommand \BibitemShut  [1]{\csname bibitem#1\endcsname}%
\let\auto@bib@innerbib\@empty
\bibitem [{\citenamefont {Van~Kampen}(1992)}]{van1992stochastic}%
  \BibitemOpen
  \bibfield  {author} {\bibinfo {author} {\bibfnamefont {N.~G.}\ \bibnamefont
  {Van~Kampen}},\ }\href@noop {} {\emph {\bibinfo {title} {Stochastic processes
  in physics and chemistry}}}\ (\bibinfo  {publisher} {Elsevier, New York},\
  \bibinfo {year} {1992})\BibitemShut {NoStop}%
\bibitem [{\citenamefont {Kessler}\ and\ \citenamefont
  {Barkai}(2010)}]{Kessler2010}%
  \BibitemOpen
  \bibfield  {author} {\bibinfo {author} {\bibfnamefont {D.~A.}\ \bibnamefont
  {Kessler}}\ and\ \bibinfo {author} {\bibfnamefont {E.}~\bibnamefont
  {Barkai}},\ }\href {\doibase 10.1103/PhysRevLett.105.120602} {\bibfield
  {journal} {\bibinfo  {journal} {Phys. Rev. Lett.}\ }\textbf {\bibinfo
  {volume} {105}},\ \bibinfo {pages} {120602} (\bibinfo {year}
  {2010})}\BibitemShut {NoStop}%
\bibitem [{\citenamefont {Lutz}\ and\ \citenamefont
  {Renzoni}(2013)}]{lutz2013}%
  \BibitemOpen
  \bibfield  {author} {\bibinfo {author} {\bibfnamefont {E.}~\bibnamefont
  {Lutz}}\ and\ \bibinfo {author} {\bibfnamefont {F.}~\bibnamefont {Renzoni}},\
  }\href@noop {} {\bibfield  {journal} {\bibinfo  {journal} {Nat. Phys.}\
  }\textbf {\bibinfo {volume} {9}},\ \bibinfo {pages} {615} (\bibinfo {year}
  {2013})}\BibitemShut {NoStop}%
\bibitem [{\citenamefont {Rebenshtok}\ \emph {et~al.}(2014)\citenamefont
  {Rebenshtok}, \citenamefont {Denisov}, \citenamefont {H\"anggi},\ and\
  \citenamefont {Barkai}}]{Rebenshtok2014}%
  \BibitemOpen
  \bibfield  {author} {\bibinfo {author} {\bibfnamefont {A.}~\bibnamefont
  {Rebenshtok}}, \bibinfo {author} {\bibfnamefont {S.}~\bibnamefont {Denisov}},
  \bibinfo {author} {\bibfnamefont {P.}~\bibnamefont {H\"anggi}}, \ and\
  \bibinfo {author} {\bibfnamefont {E.}~\bibnamefont {Barkai}},\ }\href
  {\doibase 10.1103/PhysRevLett.112.110601} {\bibfield  {journal} {\bibinfo
  {journal} {Phys. Rev. Lett.}\ }\textbf {\bibinfo {volume} {112}},\ \bibinfo
  {pages} {110601} (\bibinfo {year} {2014})}\BibitemShut {NoStop}%
\bibitem [{\citenamefont {Holz}\ \emph {et~al.}(2015)\citenamefont {Holz},
  \citenamefont {Dechant},\ and\ \citenamefont {Lutz}}]{Holz2015}%
  \BibitemOpen
  \bibfield  {author} {\bibinfo {author} {\bibfnamefont {P.~C.}\ \bibnamefont
  {Holz}}, \bibinfo {author} {\bibfnamefont {A.}~\bibnamefont {Dechant}}, \
  and\ \bibinfo {author} {\bibfnamefont {E.}~\bibnamefont {Lutz}},\ }\href
  {http://stacks.iop.org/0295-5075/109/i=2/a=23001} {\bibfield  {journal}
  {\bibinfo  {journal} {Europhys. Lett.}\ }\textbf {\bibinfo {volume} {109}},\
  \bibinfo {pages} {23001} (\bibinfo {year} {2015})}\BibitemShut {NoStop}%
\bibitem [{\citenamefont {Leibovich}\ and\ \citenamefont
  {Barkai}(2019)}]{Leibovich2019}%
  \BibitemOpen
  \bibfield  {author} {\bibinfo {author} {\bibfnamefont {N.}~\bibnamefont
  {Leibovich}}\ and\ \bibinfo {author} {\bibfnamefont {E.}~\bibnamefont
  {Barkai}},\ }\href {\doibase 10.1103/PhysRevE.99.042138} {\bibfield
  {journal} {\bibinfo  {journal} {Phys. Rev. E}\ }\textbf {\bibinfo {volume}
  {99}},\ \bibinfo {pages} {042138} (\bibinfo {year} {2019})}\BibitemShut
  {NoStop}%
\bibitem [{\citenamefont {Aghion}\ \emph {et~al.}(2019)\citenamefont {Aghion},
  \citenamefont {Kessler},\ and\ \citenamefont {Barkai}}]{Aghion2019}%
  \BibitemOpen
  \bibfield  {author} {\bibinfo {author} {\bibfnamefont {E.}~\bibnamefont
  {Aghion}}, \bibinfo {author} {\bibfnamefont {D.~A.}\ \bibnamefont {Kessler}},
  \ and\ \bibinfo {author} {\bibfnamefont {E.}~\bibnamefont {Barkai}},\ }\href
  {\doibase 10.1103/PhysRevLett.122.010601} {\bibfield  {journal} {\bibinfo
  {journal} {Phys. Rev. Lett.}\ }\textbf {\bibinfo {volume} {122}},\ \bibinfo
  {pages} {010601} (\bibinfo {year} {2019})}\BibitemShut {NoStop}%
\bibitem [{\citenamefont {Aghion}\ \emph {et~al.}(2020)\citenamefont {Aghion},
  \citenamefont {Kessler},\ and\ \citenamefont {Barkai}}]{aghion2020infinite}%
  \BibitemOpen
  \bibfield  {author} {\bibinfo {author} {\bibfnamefont {E.}~\bibnamefont
  {Aghion}}, \bibinfo {author} {\bibfnamefont {D.~A.}\ \bibnamefont {Kessler}},
  \ and\ \bibinfo {author} {\bibfnamefont {E.}~\bibnamefont {Barkai}},\
  }\href@noop {} {\bibfield  {journal} {\bibinfo  {journal} {Chaos, Solitons \&
  Fractals}\ }\textbf {\bibinfo {volume} {138}},\ \bibinfo {pages} {109890}
  (\bibinfo {year} {2020})}\BibitemShut {NoStop}%
\bibitem [{\citenamefont {Aghion}\ \emph {et~al.}(2021)\citenamefont {Aghion},
  \citenamefont {Meyer}, \citenamefont {Adlakha}, \citenamefont {Kantz},\ and\
  \citenamefont {Bassler}}]{aghion2021moses}%
  \BibitemOpen
  \bibfield  {author} {\bibinfo {author} {\bibfnamefont {E.}~\bibnamefont
  {Aghion}}, \bibinfo {author} {\bibfnamefont {P.~G.}\ \bibnamefont {Meyer}},
  \bibinfo {author} {\bibfnamefont {V.}~\bibnamefont {Adlakha}}, \bibinfo
  {author} {\bibfnamefont {H.}~\bibnamefont {Kantz}}, \ and\ \bibinfo {author}
  {\bibfnamefont {K.~E.}\ \bibnamefont {Bassler}},\ }\href@noop {} {\bibfield
  {journal} {\bibinfo  {journal} {New J. Phys.}\ }\textbf {\bibinfo {volume}
  {23}},\ \bibinfo {pages} {023002} (\bibinfo {year} {2021})}\BibitemShut
  {NoStop}%
\bibitem [{\citenamefont {Strei\ss{}nig}\ and\ \citenamefont
  {Kantz}(2021)}]{Streissnin2021}%
  \BibitemOpen
  \bibfield  {author} {\bibinfo {author} {\bibfnamefont {C.}~\bibnamefont
  {Strei\ss{}nig}}\ and\ \bibinfo {author} {\bibfnamefont {H.}~\bibnamefont
  {Kantz}},\ }\href {\doibase 10.1103/PhysRevResearch.3.013115} {\bibfield
  {journal} {\bibinfo  {journal} {Phys. Rev. Research}\ }\textbf {\bibinfo
  {volume} {3}},\ \bibinfo {pages} {013115} (\bibinfo {year}
  {2021})}\BibitemShut {NoStop}%
\bibitem [{\citenamefont {Thaler}(1983)}]{Thaler1983}%
  \BibitemOpen
  \bibfield  {author} {\bibinfo {author} {\bibfnamefont {M.}~\bibnamefont
  {Thaler}},\ }\href@noop {} {\bibfield  {journal} {\bibinfo  {journal} {Isr.
  J. Math.}\ }\textbf {\bibinfo {volume} {46}},\ \bibinfo {pages} {67}
  (\bibinfo {year} {1983})}\BibitemShut {NoStop}%
\bibitem [{\citenamefont {Aaronson}(1997)}]{Aaronson1997}%
  \BibitemOpen
  \bibfield  {author} {\bibinfo {author} {\bibfnamefont {J.}~\bibnamefont
  {Aaronson}},\ }\href@noop {} {\emph {\bibinfo {title} {An Introduction to
  Infinite Ergodic Theory}}}\ (\bibinfo  {publisher} {American Mathematical
  Society},\ \bibinfo {address} {Providence},\ \bibinfo {year}
  {1997})\BibitemShut {NoStop}%
\bibitem [{\citenamefont {Inoue}(1997)}]{inoue1997ratio}%
  \BibitemOpen
  \bibfield  {author} {\bibinfo {author} {\bibfnamefont {T.}~\bibnamefont
  {Inoue}},\ }\href@noop {} {\bibfield  {journal} {\bibinfo  {journal} {Ergod.
  Theory Dyn. Syst.}\ }\textbf {\bibinfo {volume} {17}},\ \bibinfo {pages}
  {625} (\bibinfo {year} {1997})}\BibitemShut {NoStop}%
\bibitem [{\citenamefont {Thaler}(1998)}]{Thaler1998}%
  \BibitemOpen
  \bibfield  {author} {\bibinfo {author} {\bibfnamefont {M.}~\bibnamefont
  {Thaler}},\ }\href@noop {} {\bibfield  {journal} {\bibinfo  {journal} {Trans.
  Am. Math. Soc.}\ }\textbf {\bibinfo {volume} {350}},\ \bibinfo {pages} {4593}
  (\bibinfo {year} {1998})}\BibitemShut {NoStop}%
\bibitem [{\citenamefont {Thaler}(2002)}]{Thaler2002}%
  \BibitemOpen
  \bibfield  {author} {\bibinfo {author} {\bibfnamefont {M.}~\bibnamefont
  {Thaler}},\ }\href@noop {} {\bibfield  {journal} {\bibinfo  {journal} {Ergod.
  Theory Dyn. Syst.}\ }\textbf {\bibinfo {volume} {22}},\ \bibinfo {pages}
  {1289} (\bibinfo {year} {2002})}\BibitemShut {NoStop}%
\bibitem [{\citenamefont {Inoue}(2004)}]{inoue2004ergodic}%
  \BibitemOpen
  \bibfield  {author} {\bibinfo {author} {\bibfnamefont {T.}~\bibnamefont
  {Inoue}},\ }\href@noop {} {\bibfield  {journal} {\bibinfo  {journal} {Ergod.
  Theory Dyn. Syst.}\ }\textbf {\bibinfo {volume} {24}},\ \bibinfo {pages}
  {525} (\bibinfo {year} {2004})}\BibitemShut {NoStop}%
\bibitem [{\citenamefont {Akimoto}(2008)}]{Akimoto2008}%
  \BibitemOpen
  \bibfield  {author} {\bibinfo {author} {\bibfnamefont {T.}~\bibnamefont
  {Akimoto}},\ }\href@noop {} {\bibfield  {journal} {\bibinfo  {journal} {J.
  Stat. Phys.}\ }\textbf {\bibinfo {volume} {132}},\ \bibinfo {pages} {171}
  (\bibinfo {year} {2008})}\BibitemShut {NoStop}%
\bibitem [{\citenamefont {Akimoto}\ \emph {et~al.}(2015)\citenamefont
  {Akimoto}, \citenamefont {Shinkai},\ and\ \citenamefont
  {Aizawa}}]{Akimoto2015}%
  \BibitemOpen
  \bibfield  {author} {\bibinfo {author} {\bibfnamefont {T.}~\bibnamefont
  {Akimoto}}, \bibinfo {author} {\bibfnamefont {S.}~\bibnamefont {Shinkai}}, \
  and\ \bibinfo {author} {\bibfnamefont {Y.}~\bibnamefont {Aizawa}},\
  }\href@noop {} {\bibfield  {journal} {\bibinfo  {journal} {J. Stat. Phys.}\
  }\textbf {\bibinfo {volume} {158}},\ \bibinfo {pages} {476} (\bibinfo {year}
  {2015})}\BibitemShut {NoStop}%
\bibitem [{\citenamefont {Sera}\ and\ \citenamefont {Yano}(2019)}]{Sera2019}%
  \BibitemOpen
  \bibfield  {author} {\bibinfo {author} {\bibfnamefont {T.}~\bibnamefont
  {Sera}}\ and\ \bibinfo {author} {\bibfnamefont {K.}~\bibnamefont {Yano}},\
  }\href {\doibase 10.1090/tran/7755} {\bibfield  {journal} {\bibinfo
  {journal} {Trans. Amer. Math. Soc.}\ }\textbf {\bibinfo {volume} {372}},\
  \bibinfo {pages} {3191} (\bibinfo {year} {2019})}\BibitemShut {NoStop}%
\bibitem [{\citenamefont {Sera}(2020)}]{Sera2020}%
  \BibitemOpen
  \bibfield  {author} {\bibinfo {author} {\bibfnamefont {T.}~\bibnamefont
  {Sera}},\ }\href {\doibase 10.1088/1361-6544/ab5ceb} {\bibfield  {journal}
  {\bibinfo  {journal} {Nonlinearity}\ }\textbf {\bibinfo {volume} {33}},\
  \bibinfo {pages} {1183} (\bibinfo {year} {2020})}\BibitemShut {NoStop}%
\bibitem [{\citenamefont {Aaronson}(1981)}]{Aaronson1981}%
  \BibitemOpen
  \bibfield  {author} {\bibinfo {author} {\bibfnamefont {J.}~\bibnamefont
  {Aaronson}},\ }\href@noop {} {\bibfield  {journal} {\bibinfo  {journal} {J.
  D'Analyse Math.}\ }\textbf {\bibinfo {volume} {39}},\ \bibinfo {pages} {203}
  (\bibinfo {year} {1981})}\BibitemShut {NoStop}%
\bibitem [{\citenamefont {Akimoto}\ and\ \citenamefont
  {Miyaguchi}(2010)}]{Akimoto2010}%
  \BibitemOpen
  \bibfield  {author} {\bibinfo {author} {\bibfnamefont {T.}~\bibnamefont
  {Akimoto}}\ and\ \bibinfo {author} {\bibfnamefont {T.}~\bibnamefont
  {Miyaguchi}},\ }\href@noop {} {\bibfield  {journal} {\bibinfo  {journal}
  {Phys. Rev. E}\ }\textbf {\bibinfo {volume} {82}},\ \bibinfo {pages}
  {030102(R)} (\bibinfo {year} {2010})}\BibitemShut {NoStop}%
\bibitem [{\citenamefont {Akimoto}(2012)}]{Akimoto2012}%
  \BibitemOpen
  \bibfield  {author} {\bibinfo {author} {\bibfnamefont {T.}~\bibnamefont
  {Akimoto}},\ }\href {\doibase 10.1103/PhysRevLett.108.164101} {\bibfield
  {journal} {\bibinfo  {journal} {Phys. Rev. Lett.}\ }\textbf {\bibinfo
  {volume} {108}},\ \bibinfo {pages} {164101} (\bibinfo {year}
  {2012})}\BibitemShut {NoStop}%
\bibitem [{\citenamefont {Brokmann}\ \emph {et~al.}(2003)\citenamefont
  {Brokmann}, \citenamefont {Hermier}, \citenamefont {Messin}, \citenamefont
  {Desbiolles}, \citenamefont {Bouchaud},\ and\ \citenamefont
  {Dahan}}]{Brok2003}%
  \BibitemOpen
  \bibfield  {author} {\bibinfo {author} {\bibfnamefont {X.}~\bibnamefont
  {Brokmann}}, \bibinfo {author} {\bibfnamefont {J.-P.}\ \bibnamefont
  {Hermier}}, \bibinfo {author} {\bibfnamefont {G.}~\bibnamefont {Messin}},
  \bibinfo {author} {\bibfnamefont {P.}~\bibnamefont {Desbiolles}}, \bibinfo
  {author} {\bibfnamefont {J.-P.}\ \bibnamefont {Bouchaud}}, \ and\ \bibinfo
  {author} {\bibfnamefont {M.}~\bibnamefont {Dahan}},\ }\href@noop {}
  {\bibfield  {journal} {\bibinfo  {journal} {Phys. Rev. Lett.}\ }\textbf
  {\bibinfo {volume} {90}},\ \bibinfo {pages} {120601} (\bibinfo {year}
  {2003})}\BibitemShut {NoStop}%
\bibitem [{\citenamefont {Stefani}\ \emph {et~al.}(2009)\citenamefont
  {Stefani}, \citenamefont {Hoogenboom},\ and\ \citenamefont
  {Barkai}}]{stefani2009}%
  \BibitemOpen
  \bibfield  {author} {\bibinfo {author} {\bibfnamefont {F.~D.}\ \bibnamefont
  {Stefani}}, \bibinfo {author} {\bibfnamefont {J.~P.}\ \bibnamefont
  {Hoogenboom}}, \ and\ \bibinfo {author} {\bibfnamefont {E.}~\bibnamefont
  {Barkai}},\ }\href@noop {} {\bibfield  {journal} {\bibinfo  {journal} {Phys.
  today}\ }\textbf {\bibinfo {volume} {62}},\ \bibinfo {pages} {34} (\bibinfo
  {year} {2009})}\BibitemShut {NoStop}%
\bibitem [{\citenamefont {Golding}\ and\ \citenamefont
  {Cox}(2006)}]{Golding2006}%
  \BibitemOpen
  \bibfield  {author} {\bibinfo {author} {\bibfnamefont {I.}~\bibnamefont
  {Golding}}\ and\ \bibinfo {author} {\bibfnamefont {E.~C.}\ \bibnamefont
  {Cox}},\ }\href@noop {} {\bibfield  {journal} {\bibinfo  {journal} {Phys.
  Rev. Lett.}\ }\textbf {\bibinfo {volume} {96}},\ \bibinfo {pages} {098102}
  (\bibinfo {year} {2006})}\BibitemShut {NoStop}%
\bibitem [{\citenamefont {Weigel}\ \emph {et~al.}(2011)\citenamefont {Weigel},
  \citenamefont {Simon}, \citenamefont {Tamkun},\ and\ \citenamefont
  {Krapf}}]{Weigel2011}%
  \BibitemOpen
  \bibfield  {author} {\bibinfo {author} {\bibfnamefont {A.}~\bibnamefont
  {Weigel}}, \bibinfo {author} {\bibfnamefont {B.}~\bibnamefont {Simon}},
  \bibinfo {author} {\bibfnamefont {M.}~\bibnamefont {Tamkun}}, \ and\ \bibinfo
  {author} {\bibfnamefont {D.}~\bibnamefont {Krapf}},\ }\href@noop {}
  {\bibfield  {journal} {\bibinfo  {journal} {Proc. Natl. Acad. Sci. USA}\
  }\textbf {\bibinfo {volume} {108}},\ \bibinfo {pages} {6438} (\bibinfo {year}
  {2011})}\BibitemShut {NoStop}%
\bibitem [{\citenamefont {Jeon}\ \emph {et~al.}(2011)\citenamefont {Jeon},
  \citenamefont {Tejedor}, \citenamefont {Burov}, \citenamefont {Barkai},
  \citenamefont {Selhuber-Unkel}, \citenamefont {Berg-S\o{}rensen},
  \citenamefont {Oddershede},\ and\ \citenamefont {Metzler}}]{Jeon2011}%
  \BibitemOpen
  \bibfield  {author} {\bibinfo {author} {\bibfnamefont {J.-H.}\ \bibnamefont
  {Jeon}}, \bibinfo {author} {\bibfnamefont {V.}~\bibnamefont {Tejedor}},
  \bibinfo {author} {\bibfnamefont {S.}~\bibnamefont {Burov}}, \bibinfo
  {author} {\bibfnamefont {E.}~\bibnamefont {Barkai}}, \bibinfo {author}
  {\bibfnamefont {C.}~\bibnamefont {Selhuber-Unkel}}, \bibinfo {author}
  {\bibfnamefont {K.}~\bibnamefont {Berg-S\o{}rensen}}, \bibinfo {author}
  {\bibfnamefont {L.}~\bibnamefont {Oddershede}}, \ and\ \bibinfo {author}
  {\bibfnamefont {R.}~\bibnamefont {Metzler}},\ }\href {\doibase
  10.1103/PhysRevLett.106.048103} {\bibfield  {journal} {\bibinfo  {journal}
  {Phys. Rev. Lett.}\ }\textbf {\bibinfo {volume} {106}},\ \bibinfo {pages}
  {048103} (\bibinfo {year} {2011})}\BibitemShut {NoStop}%
\bibitem [{\citenamefont {H{\"o}fling}\ and\ \citenamefont
  {Franosch}(2013)}]{Hofling2013}%
  \BibitemOpen
  \bibfield  {author} {\bibinfo {author} {\bibfnamefont {F.}~\bibnamefont
  {H{\"o}fling}}\ and\ \bibinfo {author} {\bibfnamefont {T.}~\bibnamefont
  {Franosch}},\ }\href@noop {} {\bibfield  {journal} {\bibinfo  {journal} {Rep.
  Prog. Phys.}\ }\textbf {\bibinfo {volume} {76}},\ \bibinfo {pages} {046602}
  (\bibinfo {year} {2013})}\BibitemShut {NoStop}%
\bibitem [{\citenamefont {Manzo}\ \emph {et~al.}(2015)\citenamefont {Manzo},
  \citenamefont {Torreno-Pina}, \citenamefont {Massignan}, \citenamefont
  {Lapeyre~Jr}, \citenamefont {Lewenstein},\ and\ \citenamefont
  {Parajo}}]{Manzo2015}%
  \BibitemOpen
  \bibfield  {author} {\bibinfo {author} {\bibfnamefont {C.}~\bibnamefont
  {Manzo}}, \bibinfo {author} {\bibfnamefont {J.~A.}\ \bibnamefont
  {Torreno-Pina}}, \bibinfo {author} {\bibfnamefont {P.}~\bibnamefont
  {Massignan}}, \bibinfo {author} {\bibfnamefont {G.~J.}\ \bibnamefont
  {Lapeyre~Jr}}, \bibinfo {author} {\bibfnamefont {M.}~\bibnamefont
  {Lewenstein}}, \ and\ \bibinfo {author} {\bibfnamefont {M.~F.~G.}\
  \bibnamefont {Parajo}},\ }\href@noop {} {\bibfield  {journal} {\bibinfo
  {journal} {Phys. Rev. X}\ }\textbf {\bibinfo {volume} {5}},\ \bibinfo {pages}
  {011021} (\bibinfo {year} {2015})}\BibitemShut {NoStop}%
\bibitem [{\citenamefont {Takeuchi}\ and\ \citenamefont
  {Akimoto}(2016)}]{takeuchi2016}%
  \BibitemOpen
  \bibfield  {author} {\bibinfo {author} {\bibfnamefont {K.~A.}\ \bibnamefont
  {Takeuchi}}\ and\ \bibinfo {author} {\bibfnamefont {T.}~\bibnamefont
  {Akimoto}},\ }\href@noop {} {\bibfield  {journal} {\bibinfo  {journal} {J.
  Stat. Phys.}\ }\textbf {\bibinfo {volume} {164}},\ \bibinfo {pages} {1167}
  (\bibinfo {year} {2016})}\BibitemShut {NoStop}%
\bibitem [{\citenamefont {Cohen-Tannoudji}\ and\ \citenamefont
  {Phillips}(1990)}]{cohen1990new}%
  \BibitemOpen
  \bibfield  {author} {\bibinfo {author} {\bibfnamefont {C.}~\bibnamefont
  {Cohen-Tannoudji}}\ and\ \bibinfo {author} {\bibfnamefont {W.~D.}\
  \bibnamefont {Phillips}},\ }\href@noop {} {\bibfield  {journal} {\bibinfo
  {journal} {Phys. Today}\ }\textbf {\bibinfo {volume} {43}},\ \bibinfo {pages}
  {33} (\bibinfo {year} {1990})}\BibitemShut {NoStop}%
\bibitem [{\citenamefont {Bardou}\ \emph {et~al.}(1994)\citenamefont {Bardou},
  \citenamefont {Bouchaud}, \citenamefont {Emile}, \citenamefont {Aspect},\
  and\ \citenamefont {Cohen-Tannoudji}}]{Bardou1994}%
  \BibitemOpen
  \bibfield  {author} {\bibinfo {author} {\bibfnamefont {F.}~\bibnamefont
  {Bardou}}, \bibinfo {author} {\bibfnamefont {J.~P.}\ \bibnamefont
  {Bouchaud}}, \bibinfo {author} {\bibfnamefont {O.}~\bibnamefont {Emile}},
  \bibinfo {author} {\bibfnamefont {A.}~\bibnamefont {Aspect}}, \ and\ \bibinfo
  {author} {\bibfnamefont {C.}~\bibnamefont {Cohen-Tannoudji}},\ }\href
  {\doibase 10.1103/PhysRevLett.72.203} {\bibfield  {journal} {\bibinfo
  {journal} {Phys. Rev. Lett.}\ }\textbf {\bibinfo {volume} {72}},\ \bibinfo
  {pages} {203} (\bibinfo {year} {1994})}\BibitemShut {NoStop}%
\bibitem [{\citenamefont {Barkai}\ \emph {et~al.}(2021)\citenamefont {Barkai},
  \citenamefont {Radons},\ and\ \citenamefont {Akimoto}}]{Barkai2021}%
  \BibitemOpen
  \bibfield  {author} {\bibinfo {author} {\bibfnamefont {E.}~\bibnamefont
  {Barkai}}, \bibinfo {author} {\bibfnamefont {G.}~\bibnamefont {Radons}}, \
  and\ \bibinfo {author} {\bibfnamefont {T.}~\bibnamefont {Akimoto}},\ }\href
  {\doibase 10.1103/PhysRevLett.127.140605} {\bibfield  {journal} {\bibinfo
  {journal} {Phys. Rev. Lett.}\ }\textbf {\bibinfo {volume} {127}},\ \bibinfo
  {pages} {140605} (\bibinfo {year} {2021})}\BibitemShut {NoStop}%
\bibitem [{\citenamefont {Barkai}\ \emph {et~al.}(2022)\citenamefont {Barkai},
  \citenamefont {Radons},\ and\ \citenamefont {Akimoto}}]{barkai2022gas}%
  \BibitemOpen
  \bibfield  {author} {\bibinfo {author} {\bibfnamefont {E.}~\bibnamefont
  {Barkai}}, \bibinfo {author} {\bibfnamefont {G.}~\bibnamefont {Radons}}, \
  and\ \bibinfo {author} {\bibfnamefont {T.}~\bibnamefont {Akimoto}},\
  }\href@noop {} {\bibfield  {journal} {\bibinfo  {journal} {J. Chem. Phys.}\
  }\textbf {\bibinfo {volume} {156}},\ \bibinfo {pages} {044118} (\bibinfo
  {year} {2022})}\BibitemShut {NoStop}%
\bibitem [{\citenamefont {Bardou}\ \emph {et~al.}(2002)\citenamefont {Bardou},
  \citenamefont {Bouchaud}, \citenamefont {Aspect},\ and\ \citenamefont
  {Cohen-Tannoudji}}]{Bardou2002}%
  \BibitemOpen
  \bibfield  {author} {\bibinfo {author} {\bibfnamefont {F.}~\bibnamefont
  {Bardou}}, \bibinfo {author} {\bibfnamefont {J.-P.}\ \bibnamefont
  {Bouchaud}}, \bibinfo {author} {\bibfnamefont {A.}~\bibnamefont {Aspect}}, \
  and\ \bibinfo {author} {\bibfnamefont {C.}~\bibnamefont {Cohen-Tannoudji}},\
  }\href@noop {} {\emph {\bibinfo {title} {Levy statistics and laser cooling:
  how rare events bring atoms to rest}}}\ (\bibinfo  {publisher} {Cambridge
  University Press},\ \bibinfo {year} {2002})\BibitemShut {NoStop}%
\bibitem [{\citenamefont {Aspect}\ \emph {et~al.}(1988)\citenamefont {Aspect},
  \citenamefont {Arimondo}, \citenamefont {Kaiser}, \citenamefont
  {Vansteenkiste},\ and\ \citenamefont {Cohen-Tannoudji}}]{AAK88}%
  \BibitemOpen
  \bibfield  {author} {\bibinfo {author} {\bibfnamefont {A.}~\bibnamefont
  {Aspect}}, \bibinfo {author} {\bibfnamefont {E.}~\bibnamefont {Arimondo}},
  \bibinfo {author} {\bibfnamefont {R.}~\bibnamefont {Kaiser}}, \bibinfo
  {author} {\bibfnamefont {N.}~\bibnamefont {Vansteenkiste}}, \ and\ \bibinfo
  {author} {\bibfnamefont {C.}~\bibnamefont {Cohen-Tannoudji}},\ }\href
  {\doibase 10.1103/PhysRevLett.61.826} {\bibfield  {journal} {\bibinfo
  {journal} {Phys. Rev. Lett.}\ }\textbf {\bibinfo {volume} {61}},\ \bibinfo
  {pages} {826} (\bibinfo {year} {1988})}\BibitemShut {NoStop}%
\bibitem [{\citenamefont {Kasevich}\ and\ \citenamefont {Chu}(1992)}]{KaC92}%
  \BibitemOpen
  \bibfield  {author} {\bibinfo {author} {\bibfnamefont {M.}~\bibnamefont
  {Kasevich}}\ and\ \bibinfo {author} {\bibfnamefont {S.}~\bibnamefont {Chu}},\
  }\href {\doibase 10.1103/PhysRevLett.69.1741} {\bibfield  {journal} {\bibinfo
   {journal} {Phys. Rev. Lett.}\ }\textbf {\bibinfo {volume} {69}},\ \bibinfo
  {pages} {1741} (\bibinfo {year} {1992})}\BibitemShut {NoStop}%
\bibitem [{\citenamefont {Saubam\'ea}\ \emph {et~al.}(1999)\citenamefont
  {Saubam\'ea}, \citenamefont {Leduc},\ and\ \citenamefont
  {Cohen-Tannoudji}}]{Saubamea1999}%
  \BibitemOpen
  \bibfield  {author} {\bibinfo {author} {\bibfnamefont {B.}~\bibnamefont
  {Saubam\'ea}}, \bibinfo {author} {\bibfnamefont {M.}~\bibnamefont {Leduc}}, \
  and\ \bibinfo {author} {\bibfnamefont {C.}~\bibnamefont {Cohen-Tannoudji}},\
  }\href {\doibase 10.1103/PhysRevLett.83.3796} {\bibfield  {journal} {\bibinfo
   {journal} {Phys. Rev. Lett.}\ }\textbf {\bibinfo {volume} {83}},\ \bibinfo
  {pages} {3796} (\bibinfo {year} {1999})}\BibitemShut {NoStop}%
\bibitem [{\citenamefont {Bertin}\ and\ \citenamefont
  {Bardou}(2008)}]{bertin2008}%
  \BibitemOpen
  \bibfield  {author} {\bibinfo {author} {\bibfnamefont {E.}~\bibnamefont
  {Bertin}}\ and\ \bibinfo {author} {\bibfnamefont {F.}~\bibnamefont
  {Bardou}},\ }\href@noop {} {\bibfield  {journal} {\bibinfo  {journal} {Am. J.
  Phys.}\ }\textbf {\bibinfo {volume} {76}},\ \bibinfo {pages} {630} (\bibinfo
  {year} {2008})}\BibitemShut {NoStop}%
\bibitem [{\citenamefont {Reichel}\ \emph {et~al.}(1995)\citenamefont
  {Reichel}, \citenamefont {Bardou}, \citenamefont {Dahan}, \citenamefont
  {Peik}, \citenamefont {Rand}, \citenamefont {Salomon},\ and\ \citenamefont
  {Cohen-Tannoudji}}]{RBB95}%
  \BibitemOpen
  \bibfield  {author} {\bibinfo {author} {\bibfnamefont {J.}~\bibnamefont
  {Reichel}}, \bibinfo {author} {\bibfnamefont {F.}~\bibnamefont {Bardou}},
  \bibinfo {author} {\bibfnamefont {M.~B.}\ \bibnamefont {Dahan}}, \bibinfo
  {author} {\bibfnamefont {E.}~\bibnamefont {Peik}}, \bibinfo {author}
  {\bibfnamefont {S.}~\bibnamefont {Rand}}, \bibinfo {author} {\bibfnamefont
  {C.}~\bibnamefont {Salomon}}, \ and\ \bibinfo {author} {\bibfnamefont
  {C.}~\bibnamefont {Cohen-Tannoudji}},\ }\href {\doibase
  10.1103/PhysRevLett.75.4575} {\bibfield  {journal} {\bibinfo  {journal}
  {Phys. Rev. Lett.}\ }\textbf {\bibinfo {volume} {75}},\ \bibinfo {pages}
  {4575} (\bibinfo {year} {1995})}\BibitemShut {NoStop}%
\bibitem [{\citenamefont {Akimoto}\ \emph {et~al.}(2020)\citenamefont
  {Akimoto}, \citenamefont {Barkai},\ and\ \citenamefont
  {Radons}}]{Akimoto2020}%
  \BibitemOpen
  \bibfield  {author} {\bibinfo {author} {\bibfnamefont {T.}~\bibnamefont
  {Akimoto}}, \bibinfo {author} {\bibfnamefont {E.}~\bibnamefont {Barkai}}, \
  and\ \bibinfo {author} {\bibfnamefont {G.}~\bibnamefont {Radons}},\ }\href
  {\doibase 10.1103/PhysRevE.101.052112} {\bibfield  {journal} {\bibinfo
  {journal} {Phys. Rev. E}\ }\textbf {\bibinfo {volume} {101}},\ \bibinfo
  {pages} {052112} (\bibinfo {year} {2020})}\BibitemShut {NoStop}%
\bibitem [{\citenamefont {Cox}(1962)}]{Cox1962}%
  \BibitemOpen
  \bibfield  {author} {\bibinfo {author} {\bibfnamefont {D.~R.}\ \bibnamefont
  {Cox}},\ }\href@noop {} {\emph {\bibinfo {title} {Renewal theory}}}\
  (\bibinfo  {publisher} {Methuen},\ \bibinfo {address} {London},\ \bibinfo
  {year} {1962})\BibitemShut {NoStop}%
\bibitem [{\citenamefont {Darling}\ and\ \citenamefont
  {Kac}(1957)}]{Darling1957}%
  \BibitemOpen
  \bibfield  {author} {\bibinfo {author} {\bibfnamefont {D.~A.}\ \bibnamefont
  {Darling}}\ and\ \bibinfo {author} {\bibfnamefont {M.}~\bibnamefont {Kac}},\
  }\href@noop {} {\bibfield  {journal} {\bibinfo  {journal} {Trans. Am. Math.
  Soc.}\ }\textbf {\bibinfo {volume} {84}},\ \bibinfo {pages} {444} (\bibinfo
  {year} {1957})}\BibitemShut {NoStop}%
\bibitem [{\citenamefont {Shinkai}\ and\ \citenamefont
  {Aizawa}(2006)}]{shinkai2006lempel}%
  \BibitemOpen
  \bibfield  {author} {\bibinfo {author} {\bibfnamefont {S.}~\bibnamefont
  {Shinkai}}\ and\ \bibinfo {author} {\bibfnamefont {Y.}~\bibnamefont
  {Aizawa}},\ }\href@noop {} {\bibfield  {journal} {\bibinfo  {journal} {Prog.
  Theor. Phys.}\ }\textbf {\bibinfo {volume} {116}},\ \bibinfo {pages} {503}
  (\bibinfo {year} {2006})}\BibitemShut {NoStop}%
\bibitem [{\citenamefont {Kasahara}(1977)}]{kasahara77}%
  \BibitemOpen
  \bibfield  {author} {\bibinfo {author} {\bibfnamefont {Y.}~\bibnamefont
  {Kasahara}},\ }\href@noop {} {\bibfield  {journal} {\bibinfo  {journal}
  {Publ. RIMS, Kyoto Univ.}\ }\textbf {\bibinfo {volume} {12}},\ \bibinfo
  {pages} {801} (\bibinfo {year} {1977})}\BibitemShut {NoStop}%
\bibitem [{\citenamefont {Lubelski}\ \emph {et~al.}(2008)\citenamefont
  {Lubelski}, \citenamefont {Sokolov},\ and\ \citenamefont
  {Klafter}}]{Lubelski2008}%
  \BibitemOpen
  \bibfield  {author} {\bibinfo {author} {\bibfnamefont {A.}~\bibnamefont
  {Lubelski}}, \bibinfo {author} {\bibfnamefont {I.~M.}\ \bibnamefont
  {Sokolov}}, \ and\ \bibinfo {author} {\bibfnamefont {J.}~\bibnamefont
  {Klafter}},\ }\href@noop {} {\bibfield  {journal} {\bibinfo  {journal} {Phys.
  Rev. Lett.}\ }\textbf {\bibinfo {volume} {100}},\ \bibinfo {pages} {250602}
  (\bibinfo {year} {2008})}\BibitemShut {NoStop}%
\bibitem [{\citenamefont {He}\ \emph {et~al.}(2008)\citenamefont {He},
  \citenamefont {Burov}, \citenamefont {Metzler},\ and\ \citenamefont
  {Barkai}}]{He2008}%
  \BibitemOpen
  \bibfield  {author} {\bibinfo {author} {\bibfnamefont {Y.}~\bibnamefont
  {He}}, \bibinfo {author} {\bibfnamefont {S.}~\bibnamefont {Burov}}, \bibinfo
  {author} {\bibfnamefont {R.}~\bibnamefont {Metzler}}, \ and\ \bibinfo
  {author} {\bibfnamefont {E.}~\bibnamefont {Barkai}},\ }\href@noop {}
  {\bibfield  {journal} {\bibinfo  {journal} {Phys. Rev. Lett.}\ }\textbf
  {\bibinfo {volume} {101}},\ \bibinfo {pages} {058101} (\bibinfo {year}
  {2008})}\BibitemShut {NoStop}%
\bibitem [{\citenamefont {Miyaguchi}\ and\ \citenamefont
  {Akimoto}(2011)}]{Miyaguchi2011}%
  \BibitemOpen
  \bibfield  {author} {\bibinfo {author} {\bibfnamefont {T.}~\bibnamefont
  {Miyaguchi}}\ and\ \bibinfo {author} {\bibfnamefont {T.}~\bibnamefont
  {Akimoto}},\ }\href@noop {} {\bibfield  {journal} {\bibinfo  {journal} {Phys.
  Rev. E}\ }\textbf {\bibinfo {volume} {83}},\ \bibinfo {pages} {031926}
  (\bibinfo {year} {2011})}\BibitemShut {NoStop}%
\bibitem [{\citenamefont {Miyaguchi}\ and\ \citenamefont
  {Akimoto}(2013)}]{Miyaguchi2013}%
  \BibitemOpen
  \bibfield  {author} {\bibinfo {author} {\bibfnamefont {T.}~\bibnamefont
  {Miyaguchi}}\ and\ \bibinfo {author} {\bibfnamefont {T.}~\bibnamefont
  {Akimoto}},\ }\href {\doibase 10.1103/PhysRevE.87.032130} {\bibfield
  {journal} {\bibinfo  {journal} {Phys. Rev. E}\ }\textbf {\bibinfo {volume}
  {87}},\ \bibinfo {pages} {032130} (\bibinfo {year} {2013})}\BibitemShut
  {NoStop}%
\bibitem [{\citenamefont {Akimoto}\ and\ \citenamefont
  {Miyaguchi}(2013)}]{Akimoto2013a}%
  \BibitemOpen
  \bibfield  {author} {\bibinfo {author} {\bibfnamefont {T.}~\bibnamefont
  {Akimoto}}\ and\ \bibinfo {author} {\bibfnamefont {T.}~\bibnamefont
  {Miyaguchi}},\ }\href@noop {} {\bibfield  {journal} {\bibinfo  {journal}
  {Phys. Rev. E}\ }\textbf {\bibinfo {volume} {87}},\ \bibinfo {pages} {062134}
  (\bibinfo {year} {2013})}\BibitemShut {NoStop}%
\bibitem [{\citenamefont {Akimoto}\ and\ \citenamefont
  {Yamamoto}(2016)}]{AkimotoYamamoto2016a}%
  \BibitemOpen
  \bibfield  {author} {\bibinfo {author} {\bibfnamefont {T.}~\bibnamefont
  {Akimoto}}\ and\ \bibinfo {author} {\bibfnamefont {E.}~\bibnamefont
  {Yamamoto}},\ }\href@noop {} {\bibfield  {journal} {\bibinfo  {journal} {J.
  Stat. Mech.}\ }\textbf {\bibinfo {volume} {2016}},\ \bibinfo {pages} {123201}
  (\bibinfo {year} {2016})}\BibitemShut {NoStop}%
\bibitem [{\citenamefont {Albers}\ and\ \citenamefont
  {Radons}(2018)}]{Albers2018}%
  \BibitemOpen
  \bibfield  {author} {\bibinfo {author} {\bibfnamefont {T.}~\bibnamefont
  {Albers}}\ and\ \bibinfo {author} {\bibfnamefont {G.}~\bibnamefont
  {Radons}},\ }\href {\doibase 10.1103/PhysRevLett.120.104501} {\bibfield
  {journal} {\bibinfo  {journal} {Phys. Rev. Lett.}\ }\textbf {\bibinfo
  {volume} {120}},\ \bibinfo {pages} {104501} (\bibinfo {year}
  {2018})}\BibitemShut {NoStop}%
\bibitem [{\citenamefont {Radice}\ \emph {et~al.}(2020)\citenamefont {Radice},
  \citenamefont {Onofri}, \citenamefont {Artuso},\ and\ \citenamefont
  {Pozzoli}}]{Radice2020}%
  \BibitemOpen
  \bibfield  {author} {\bibinfo {author} {\bibfnamefont {M.}~\bibnamefont
  {Radice}}, \bibinfo {author} {\bibfnamefont {M.}~\bibnamefont {Onofri}},
  \bibinfo {author} {\bibfnamefont {R.}~\bibnamefont {Artuso}}, \ and\ \bibinfo
  {author} {\bibfnamefont {G.}~\bibnamefont {Pozzoli}},\ }\href {\doibase
  10.1103/PhysRevE.101.042103} {\bibfield  {journal} {\bibinfo  {journal}
  {Phys. Rev. E}\ }\textbf {\bibinfo {volume} {101}},\ \bibinfo {pages}
  {042103} (\bibinfo {year} {2020})}\BibitemShut {NoStop}%
\bibitem [{\citenamefont {Albers}\ and\ \citenamefont
  {Radons}(2022)}]{Albers2022}%
  \BibitemOpen
  \bibfield  {author} {\bibinfo {author} {\bibfnamefont {T.}~\bibnamefont
  {Albers}}\ and\ \bibinfo {author} {\bibfnamefont {G.}~\bibnamefont
  {Radons}},\ }\href {\doibase 10.1103/PhysRevE.105.014113} {\bibfield
  {journal} {\bibinfo  {journal} {Phys. Rev. E}\ }\textbf {\bibinfo {volume}
  {105}},\ \bibinfo {pages} {014113} (\bibinfo {year} {2022})}\BibitemShut
  {NoStop}%
\bibitem [{\citenamefont {Bouchaud}\ and\ \citenamefont
  {Georges}(1990)}]{bouchaud90}%
  \BibitemOpen
  \bibfield  {author} {\bibinfo {author} {\bibfnamefont {J.}~\bibnamefont
  {Bouchaud}}\ and\ \bibinfo {author} {\bibfnamefont {A.}~\bibnamefont
  {Georges}},\ }\href@noop {} {\bibfield  {journal} {\bibinfo  {journal} {Phys.
  Rep.}\ }\textbf {\bibinfo {volume} {195}},\ \bibinfo {pages} {127} (\bibinfo
  {year} {1990})}\BibitemShut {NoStop}%
\bibitem [{\citenamefont {Machta}(1985)}]{Machta1985}%
  \BibitemOpen
  \bibfield  {author} {\bibinfo {author} {\bibfnamefont {J.}~\bibnamefont
  {Machta}},\ }\href@noop {} {\bibfield  {journal} {\bibinfo  {journal}
  {Journal of Physics A: Mathematical and General}\ }\textbf {\bibinfo {volume}
  {18}},\ \bibinfo {pages} {L531} (\bibinfo {year} {1985})}\BibitemShut
  {NoStop}%
\bibitem [{\citenamefont {Box}\ and\ \citenamefont
  {Muller}(1958)}]{box1958note}%
  \BibitemOpen
  \bibfield  {author} {\bibinfo {author} {\bibfnamefont {G.~E.}\ \bibnamefont
  {Box}}\ and\ \bibinfo {author} {\bibfnamefont {M.~E.}\ \bibnamefont
  {Muller}},\ }\href@noop {} {\bibfield  {journal} {\bibinfo  {journal} {Ann.
  Math. Statist.}\ }\textbf {\bibinfo {volume} {29}},\ \bibinfo {pages} {610}
  (\bibinfo {year} {1958})}\BibitemShut {NoStop}%
\end{thebibliography}
%

\end{document}